\documentclass[11pt, a4paper]{article}
\usepackage{hyperref}
\usepackage[hscale=0.75,vscale=0.8]{geometry}
\usepackage{amsmath, amsthm, amssymb, mathrsfs}
\usepackage{authblk}
\usepackage{enumerate}
\usepackage{mathtools}
\newtheorem{thm}{Theorem}

\newtheorem{rem}[thm]{Remark}
\newtheorem{conj}[thm]{Conjecture}
\theoremstyle{definition}
\newtheorem{defn}[thm]{Definition}

\newcommand{\x}{{\rm x}}

\usepackage{color}
\usepackage[dvipsnames]{xcolor}

\begin{document}

\title{On the initial value problem for semiclassical gravity without and with quantum state collapses}

\author{Benito A. Ju\'arez-Aubry$^{1}$\thanks{{\tt benito.juarez@correo.nucleares.unam.mx}}, Bernard S. Kay$^{2}$\thanks{{\tt bernard.kay@york.ac.uk}}, Tonatiuh Miramontes$^{1}$\thanks{{\tt tonatiuh.miramontes@correo.nucleares.unam.mx}} \\ and Daniel Sudarsky$^{1}$\thanks{\tt sudarsky@nucleares.unam.mx}}
\affil{$^{1}$Departamento de Gravitaci\'on y Teor\'ia de Campos, Instituto de Ciencias Nucleares, Universidad Nacional Aut\'onoma de M\'exico, A. Postal 70-543, Mexico City 045010, Mexico}
\affil{$^{2}$Department of Mathematics, University of York, York YO10 5DD, UK}

\date{\today}

\maketitle

\begin{abstract}
Semiclassical gravity is the theory in which the classical Einstein tensor of a spacetime is coupled to quantum matter fields propagating on the spacetime via the expectation value of their renormalized stress-energy tensor in a quantum state.  We explore two issues, taking the Klein Gordon equation as our model quantum field theory.  The first is the provision of a suitable initial value formulation for the theory.   Towards this, we address the question, for given initial data consisting of the classical metric and its first three `time' derivatives off the surface together with a choice of initial quantum state, of what is an appropriate `surface Hadamard' condition such that, for initial data for which it is satisfied it is reasonable to conjecture that there will be a Cauchy development whose quantum state is Hadamard.  This requires dealing with the fact that, given two points on an initial surface, the spacetime geodesic between them does not, in general, lie in that surface. So the (squared) geodesic distance that occurs in the Hadamard subtraction differs from that intrinsic to the initial  surface.  We handle this complication by expanding the former as a suitable 3-dimensional covariant Taylor expansion in the latter.  Moreover the renormalized expectation value of the stress-energy tensor in the initial surface depends explicitly on the fourth, `time', derivative of the metric, which is not part of the initial data, but which we argue is given, implicitly, by the semiclassical Einstein equations on the initial surface. (The r\^ole played by those equations also entails that the surface Hadamard condition subsumes the constraints.) We also introduce the notion of \textit{physical solutions}, which, inspired by a 1993 proposal of Parker and Simon, we define to be solutions which are smooth in $\hbar$ at $\hbar=0$. We conjecture that for these solutions the second and third time derivatives of the metric will be determined 
once the first and second time derivatives are specified. We point out that a simpler treatment of the initial value problem can be had if we adopt yet more of the spirit of Parker and Simon and content ourselves with solutions to order $\hbar$ which are Hadamard to order $\hbar$.  A further simplification occurs if we consider semiclassical gravity to order $\hbar^0$.   This resembles classical general relativity in that it is free from the complications of higher derivative terms and does not require any Hadamard condition.  But it can still incorporate nontrivial quantum features such as superpositions of classical-like quantum states  of  the matter fields.  Our second issue concerns the prospects for combining semiclassical gravity with theories of spontaneous quantum state collapse.   We will focus our attention on proposals involving abrupt changes in the quantum field state which occur on certain (random, non-intersecting) Cauchy surfaces according to some -- yet to be  developed -- generally covariant \emph{objective collapse} model but that, in between such collapse surfaces, we have a physical semiclassical solution (or a solution of order $O(\hbar)$ or a solution of order $O(\hbar^0)$).  On each collapse surface, the semiclassical gravity equations will necessarily be violated and, as Page and Geilker pointed out in 1981, there will therefore necessarily be a discontinuity in the expectation value of the renormalized stress-energy tensor.  Nevertheless, we argue, based on our conjecture about the well-posedness of the initial value problem for physical solutions, that, with a suitable rule for the jump in the metric and/or the extrinsic curvature, the time evolution will still be uniquely determined. We tentatively argue that a natural jump rule would be one in which the metric itself and the transverse traceless part of the extrinsic curvature will be continuous and the jump will be confined to the remaining parts of the extrinsic curvature.  We aid and complement our discussion by studying our two issues also in the simpler cases of a semiclassical scalars model and semiclassical electrodynamics.

\end{abstract} 

\section{Introduction}

Semiclassical gravity (without state collapses) is a theory in which the gravitational field is taken to be described by a classical spacetime $({\cal M}, g_{ab})$, while matter is described by a quantum field theory on that spacetime. The dynamics is determined by demanding that, in addition to the equations of motion for the quantum field theory on the spacetime, the semiclassical Einstein equations
\begin{equation}
\label{semi-simple}
G_{ab} + \Lambda g_{ab} = 8\pi G_{\rm N} \omega(T^{\rm ren}_{ab}),
\end{equation} 
must hold, where $\omega(T_{ab}^{\rm ren})$ is the expectation value of the matter's (unsmeared) renormalized quantum stress-energy tensor in the algebraic state $\omega$ of the quantum field.  (Below, we shall sometimes refer to $\omega(T_{ab}^{\rm ren})$ simply as `the expectation value of the stress-energy tensor' or even as just `the stress-energy tensor'.)

We will take as a simple model for matter a real scalar field satisfying the Klein Gordon equation
\begin{equation}
\label{KG}
(\Box - m^2 - \xi R)\phi = 0
\end{equation}
where we include a nonminimal coupling $\xi R$ term for a constant, $\xi$.

There are at least two viewpoints that one can adopt \emph{vis-\`a-vis} semiclassical gravity. The mainstream view (which we share) is that it can, at best, be an approximation to a fundamental quantum gravity theory.  However some authors have taken the view that the quantisation of gravity is not a necessity and that semiclassical gravity can play a fundamental 
r\^ole in its own right. 

There are also a number of possible variants of the equations \eqref{semi-simple}.   Thus on the one hand, one can replace the expectation value on the right hand side of \eqref{semi-simple} with an `in-out' matrix element.   Also one can envisage terms involving  {\it gravitons} as being included with `matter' on the right hand side of \eqref{semi-simple}, or not.   Such graviton terms traditionally arise when one views semiclassical gravity as arising from the zero and one-loop terms in a loop expansion of traditional perturbative quantum gravity.   However if one e.g.\ took the point of view that there are $N$ matter fields and that the right hand side of \eqref{semi-simple} corresponds to the first term in a $1/N$ expansion, then graviton terms would be absent.   

In any case, in one or other of these variants, semiclassical gravity continues to be of great interest, and has been applied to problems or inspired studies in black hole physics \cite{Hawking:1976ra, Hawking:1992cc, Modak:2017yth, Wald, Meda:2021zdw}, cosmology \cite{Canate:2018wtx,Dappiaggi:2008mm, DiezTejedor:2011ci, Eltzner:2010nx, Gottschalk:2018kqt, Juarez-Aubry:2019jbw, Meda:2020smb, Piccirilli:2017mto, Pinamonti:2010is, Pinamonti:2013wya}, cosmic censorship \cite{Casals:2016odj, Casals:2019jfo, Dias:2019ery,  Emparan:2020rnp, Hollands:2019whz, Hollands:2020qpe, Juarez-Aubry:2015dla, Juarez-Aubry:2021tae}, quantum inequalities and singularity theorems \cite{Flanagan:1996gw, Fewster:1999gj, Fewster:2010gm, Fewster:2012yh, Fewster:2018pey, Fewster:2018pey, Ford:1994bj, Ford:1997fa}, the ruling-out of faster-than-light communication and time travel \cite{Cramer:1996vz, Cramer:1997pn, Finazzi:2009jb, Ford:1995wg, Hiscock:1997ya, Kay:2000fi, Kay:1996hj}, etc. Conceptual questions have been addressed, for instance, in \cite{Modak:2017yth, Kuo:1993if, Parker:1993dk, Struyve:2017mff}, while the mathematical structure of the theory has been studied e.g.\ in \cite{DiezTejedor:2011ci, Gottschalk:2018kqt, Meda:2020smb, Pinamonti:2010is, Pinamonti:2013wya, Flanagan:1996gw, Parker:1993dk, Juarez-Aubry:2020uim, Juarez-Aubry:2021abq, Sanders:2020osl, Ford:1984hs, Juarez-Aubry:2019jon,  Pinamonti:2013zba, Simon:1990jn, Meda:2022hdh}. Furthermore, some of the most important open questions of fundamental physics are framed in the semiclassical framework, such as the cosmological constant problem \cite{Juarez-Aubry:2019jbw, Martin:2012bt, Padilla:2015aaa, Weinberg:1988cp} or the evaporation of black holes due to Hawking radiation \cite{Hawking:1976ra, Wald, Hawking:1974sw}. A 2005 review covering many aspects of semiclassical gravity can be found in \cite{Ford}.

(Above, we have included citations to some papers which, strictly, are about quantum field theory in a fixed spacetime, but where physical inferences are made which assume the expectation value of the renormalized stress energy tensor acts as a source in the semiclassical Einstein equations.)

The present paper has two main purposes:  The first is to explore whether or how semiclassical gravity may be furnished with a suitable initial value formulation. The second, which will build on the results of the first, is to explore whether or how semiclassical gravity may be formulated so as to be consistent with quantum state collapses  in the matter sector of the theory.

Regarding the second of these purposes, we have in mind state collapses which occur stochastically and that are independent of a notion of observer, such as in so-called \emph{objective collapse} or \emph{dynamical reduction models}.  We will not consider here state reductions that, according to the Copenhagen interpretation of quantum mechanics,  occur `upon measurement', as these would  involve not only the standard conceptual difficulties, but also require that one incorporate the measuring device and its corresponding contributions to the stress-energy tensor into the description of the problem. Moreover, as discussed in \cite{Maudlin:2019} the issue of conservation of the stress-energy tensor, seems to become simply intractable in the Copenhagen interpretation of quantum theory due to its vagueness.

In taking this view, we consider that semiclassical gravity with such stochastic collapses represents an effective description of the gravity-matter interface, which would be valid as long as scales of curvature and matter densities are well below the Planck scale, and that there is an underlying theory of quantum gravity with an appropriate mechanism of quantum state reduction on whose precise nature we remain agnostic. 

Our aim will be to explore whether semiclassical gravity can be consistent with the sort of abrupt changes of state that one expects to occur on such a theory.   We shall assume that, in a semiclassical description, such abrupt changes of state occur instantaneously on certain randomly chosen non-intersecting spacelike hypersurfaces, in fact Cauchy surfaces (which we call below \emph{collapse surfaces}).  We shall not concern ourselves with the details of how these surfaces are selected, nor with how the abrupt changes of state that occur at them are determined.   As we shall discuss further below, what we shall discuss is the `jump' in the Cauchy data for the classical gravitational field that accompanies such abrupt changes in the state as we discuss further below. 

Let us say a few words about dynamical reduction models. In non-relativistic quantum mechanics, there are well known models such as GRW and CSL \cite{GRW, CSL} (see also \cite{BellJumps}) of  spontaneous collapse dynamics.   The interested reader is referred to the review 
\cite{Bassi:2003gd} or to the concise discussion in the introduction of \cite{Juarez-Aubry:2017} (an expanded discussion of which appears in the arXiv v1 of \cite{Juarez-Aubry:2017}). However, no fully-workable generally covariant models exist in the literature.
Relevant  early studies  such as   \cite{Ghirardi:1990,  Myrvold:2002,  Myrvold:2003}  helped  pave the way to some interesting developments in that direction.    See  \cite{Bedingham:2011, Pearle:2015} for proposals for how to generalize collapse models to relativistic quantum field theories.   See also  \cite{Juarez-Aubry:2017}  which, in the context of quantum field theory in a fixed curved spacetime, studies general requirements on the collapse generating operators ensuring, when assuming  that  collapses   occur on nonintersecting  Cauchy surfaces,  Hadamard states  would  collapse to Hadamard states. This feature is expected to  play a central  r\^ole  in the search for a  completely satisfactory  generally covariant collapse model  applicable  to  the context of quantum field theory in curved spacetime or to the context of semiclassical gravity.  

For our purposes here, we shall just take as an assumption that a generally covariant collapse rule can be formulated in the context of semiclassical gravity and that it will have the property that collapses take place on non-intersecting Cauchy surfaces and map `surface Hadamard states' (to be discussed below) to `surface Hadamard states'.  However we shall not need to make any assumptions here about the details of  the (stochastic) rules governing the choice of surfaces on which the collapses take place and governing the way that the states change when collapses happen; rather we shall just assume that such a theory can be formulated, and explore a fundamental   aspect of how such a theory could be  made compatible with semiclassical gravity.   Namely, the question, which we shall discuss further below, of how the classical initial data consisting of the metric and suitable time-derivatives of the metric on a collapse surface must change (or `jump') when a quantum state collapse occurs on that surface.

Turning to the first purpose of the paper, the initial value problem for semiclassical gravity (without collapses) appears to have only been addressed hitherto for certain particular spacetimes such as Friedman Lema\^itre Robertson Walker spacetimes (FLRW) 
\cite{Dappiaggi:2008mm, Meda:2020smb, Pinamonti:2010is, Pinamonti:2013wya, GottRothSiem}, static spacetimes \cite{Juarez-Aubry:2020uim, Sanders:2020osl} or special cases with conformal coupling in conformally static spacetimes \cite{Juarez-Aubry:2021abq}, while in the present paper we aim to initiate a study of the generic case.  Amongst our motivations is the hope to gain a clearer understanding of the set of solutions to semiclassical gravity.   Indeed, were we able to show that the initial value problem, when suitably stated, were well posed, that would mean that each set of allowable initial data determines a solution in much the same way that, thanks to the well-posedness (see  e.g.\ \cite[Chapter 10]{Wald}) of the initial value problem for classical general relativity (for some given reasonable matter model), we know that the specification of a 3-metric, $h_{ab}$ and a symmetric tensor $K_{ab}$ (together with suitable data for the classical matter) satisfying the constraint equations on a 3-manifold, $\cal C$ -- which we think of as the 3-dimensional surface ${0} \times {\cal C}$ in the 4-manifold ${\cal M} =\mathbb{R}\times {\cal C}$ -- determines a (globally hyperbolic) solution, $({\cal M}, g_{ab})$ to the classical field equations with the property that $h_{ab}$ is the induced metric on $\cal C$ while $K_{ab}$ is its extrinsic curvature (and the matter data on the surface arises by restriction to the surface of suitable matter variables).   Of course the value of any such result depends on how well we understand the set of allowable initial data, which in the case of general relativity, means how well we understand the set of solutions to the constraint equations. 

Aside from this motivation, an understanding of the initial value problem for semiclassical gravity will be crucial for our second purpose since we expect that an abrupt change in the quantum state on a Cauchy surface will necessarily (in view of the need to keep satisfying the constraint equations) be accompanied by a corresponding abrupt change, or `jump', in the Cauchy data for the classical gravitational field at that surface.  Indeed, an important aspect of the question of whether semiclassical gravity can be made consistent with an abrupt change of quantum state on a Cauchy surface is the question whether a precise rule can be formulated for how that jump in the Cauchy data for the classical gravitational field at that surface is determined by the change in the quantum state and we shall take some steps towards answering the question of what such a rule could be.  

To aid our discussion of both the initial value problem and of this \textit{what jumps} question we shall find it useful to discuss two simpler semiclassical theories in Minkowski space.   The first is the \textit{semiclassical scalar Yukawa model}, which has been studied previously in \cite{Juarez-Aubry:2019jon},
\begin{subequations}
\label{ScalScalar}
\begin{align}
(\partial_a\partial^a - M^2)\chi + \mu = \lambda\omega(\phi^2), \label{psieq}\\
(\partial_a\partial^a - m^2 - 2\lambda \chi)\phi =0, \label{phieq}
\end{align}
\end{subequations}
for a classical field, $\chi$, interacting with a, say Hermitian, quantum field, $\phi$.  Here, $\omega(\phi^2)$ denotes the expectation value in some quantum state, $\omega$, of the square
of the quantum $\phi$ field (which will need suitably renormalizing -- see Section \ref{Sect:SSC}).  
$\lambda$ is a coupling constant and $\mu$ another constant needed for the model to be renormalizable in a suitable sense.\footnote{In \cite{Juarez-Aubry:2019jon} the $\mu$ constant has been included in the renormalisation ambiguities of $\omega(\phi^2)$.}  (Again, see Section \ref{Sect:SSC}.) 

The second is semiclassical electrodynamics for a (now charged) scalar field, which we take to be governed by the equations in flat spacetime
\begin{subequations}
\label{SemiElectro}
\begin{align}
\partial^aF_{ab} = -4\pi \omega(j_b^{\rm ren})\label{Fabeq}\\
(\eta^{ab}(\partial_a - ieA_a)(\partial_b - ieA_b)-m^2)\phi =0,\label{chphieq}
\end{align}
\end{subequations}
where ${\omega(j_b^{\rm ren})}$ denotes the renormalized  expectation value in the quantum state, $\omega$, 
of the electric current, 
\begin{equation}
\label{ja}
j_a=-ie\left(\phi^*(\partial_a - ie A_a)\phi  - ((\partial_a - ie A_a)\phi)^*\phi\right)
\end{equation}
and where the field strength, $F_{ab}$, which is treated as classical, can be written
in the form 
$F_{ab}= \partial_a A_b - \partial_b A_a$, 
 for a vector potential, $A_a$, and the $A_a$ in Equations \eqref{chphieq} and \eqref{ja} is one such choice.

When we study the initial value problem for these models, we shall confine our interest to initial surfaces which are constant-time surfaces in some Lorentz frame and when we study the `what jumps' question, we shall assume that collapses happen at certain random times again in a single Lorentz frame.  

We are still far from having a full mathematical formulation of the initial value problem for semiclassical gravity, but, as a first step towards such an enterprise, we will attempt to state with enough precision the broad features that we expect such a formulation to possess and we will motivate, and make, a number of conjectures about what we expect to be true and be provable and also expose some open problems deserving of further study.  One of the things we need to deal with is the fact that the semiclassical gravity equations are, as we shall recall, actually fourth order (as well as nonlocal) in nature.   As is well known this presumably leads to physically spurious \emph{runaway} solutions.  (But see however \cite{GibbonsPopeSolod}.)    A proposal was made to eliminate these by Parker and Simon \cite{Parker:1993dk, Simon:1990ic} (see also \cite{Flanagan:1996gw}) by viewing $\hbar$ as a parameter on which each solution depends and restricting attention to a subset of so-called \emph{physical solutions} which they defined to be those which arise as order $\hbar$ corrections to the classical Einstein equations which result from Equation \eqref{semi-simple} (with the possible addition of a classical stress-energy term) when $\hbar$ is set to zero.  The Parker-Simon definition of `physical solution' would also seem well suited for an approach to the initial value problem in terms of perturbative solutions.

In the present paper, we modify the criterion so as to also apply to strict solutions by demanding that, to be regarded as physical, a solution should (in any coordinate chart) be jointly smooth in spacetime coordinates and in $\hbar$ at $\hbar=0$.  We shall conjecture (in Conjecture \ref{Conj:Phys}) that, by adopting this criterion for physical solutions, on a Cauchy surface, the usual two pieces of initial gravitational data, i.e.\ the 3-metric and the extrinsic curvature alone will (together with suitable initial data for the quantum state-- see below for a detailed discussion of the notion of `physical surface Hadamard state') be sufficient for a well-posed initial value problem for Equation \eqref{semi-simple} and that such data will, in fact, determine the two further pieces of Cauchy data (i.e.\ the second and third derivatives of the metric off the Cauchy 
surface\footnote{\label{Gaussian} For definiteness, for a given semiclassical solution, we may, e.g.\  define what we mean by derivatives of the metric `off' the Cauchy surface as time derivatives of the metric when it is written in Gaussian normal coordinates \cite{Wald} adapted to that surface in the spacetime of that classical solution.  When we refer, in the paper, to `time derivatives', unless otherwise stated, we do so in this sense.  Let us also note here that we use the notation $\dot g_{ij}$, $\ddot g_{ij}$ etc.\ to denote the first few time derivatives of the metric, but we shall also use the notation $g^{(3)}_{ij}$, $g^{(n)}_{ij}$  etc.\ to denote the 3rd, respectively $n$th derivative etc.} that one would expect to need to be specified in view of the 4th order character of the equations.    

An important remark here is that Parker and Simon actually only sought approximate solutions -- which solve Equation \eqref{semi-simple} up to order\footnote{Actually Parker and Simon confined their discussion to the conformally coupled massless Klein Gordon equation.  In generalizing their work to include nonzero mass here, it is convenient to treat the mass term as if it were independent of $\hbar$ and to also treat Newton's constant, $G_N$, and the cosmological constant, $\Lambda$, 
as if they are independent of $\hbar$. See Sections \ref{Sect:Dephbar} and 
\ref{Sect:1andhbar}.} $\hbar$ but may fail to solve it to order $\hbar^2$ on the grounds that, when we think of semiclassical gravity as an approximation to a fully quantum theory, quantum gravitational corrections would anyway alter the theory at order $\hbar^2$. This is a reasonable viewpoint, but in fact leaves out interesting solutions, such as Starobinsky inflation, as pointed out by Simon himself in \cite{Simon:1991bm}. The Parker-Simon criterion also leaves out solutions within the validity regime of semiclassical gravity that could have non-perturbative effects that are not captured by the first order approach, as argued in \cite{Flanagan:1996gw}.
 
Nevertheless, this does not mean that semiclassical gravity to order $\hbar$ is not relevant in some situations of physical interest. Hence, an important feature of the paper will be a discussion of the initial value problem for such approximate solutions to order $\hbar$, also since it seems to offer important simplifications and for future work we advocate using it.   We shall also advocate the use of an even simpler version of semiclassical gravity -- semiclassical gravity to order $\hbar^0$ -- either as a valuable approximate theory in its own right or as a first step towards a perturbative discussion of the theory to order $\hbar$.   However, we shall first aim to say as much as we can about the theory of general solutions or physical solutions which (in the absence of collapses) exactly solve \eqref{semi-simple} (and which, in the case of physical solutions, satisfy the above smoothness condition in $\hbar$) partly because this will help us to make clear, later, just how and in what way, the order $\hbar^0$ and order $\hbar$ approximations to the theory are simpler. 

Restricting to physical solutions will still allow a rich class of solutions, even in the case of spacetimes and matter e.g.\ with FLRW symmetry, in spite of the fact that, here, we will not contemplate the inclusion of a classical piece in the stress-energy tensor, provided we include consideration of quantum states other than just the class of quasi-free states with vanishing one-point functions.  In fact, we will contemplate states (not in that class)  which will provide us with a piece of the expectation value of the stress-energy tensor on the right hand side of \eqref{semi-simple} which tends to a nonzero value as $\hbar$ tends to zero, in addition to a piece which vanishes as $\hbar\rightarrow 0$.  For such a state, one might be tempted to denote the latter piece (which vanishes in that limit) as the `quantum part' of the stress-energy tensor and the former piece (which tends to a nonzero value) as the `classical part' and it will play a similar role to a classical piece in permitting that rich class.\footnote{\label{FRLWBirk} Say if the cosmological constant vanishes (to order $\hbar^0$) were we to restrict to FLRW symmetry and the state, $\omega$, to have a two-point function of order $\hbar$, then, since the stress-energy tensor will tend to zero as $\hbar$ tends to zero, then, by Birkhoff's theorem, restricting to physical solutions would only permit quantum corrections to Minkowski space.  Note that examples of such states are `coherent states' or, more precisely, quasifree states with nonvanishing one-point functions because, for such states, the \emph{two-point function} will have a piece of order $\hbar^0$ -- see Footnote \ref{twopoint} and Equation \eqref{trunc}.} 

But note that, as we shall discuss towards the end of Section \ref{Sect:star}, despite remaining nonzero in the $\hbar\rightarrow 0$ limit, for certain physically interesting such states -- an example is states which can be regarded as superpositions of coherent states -- the former piece can include terms which depend nontrivially on $\hbar$, e.g.\ in the just-mentioned examples, there is a term of form $A\exp(-B/\hbar)$, where $A$ and $B$ are independent of $\hbar$, and $B$ positive, and thus thinking of the former piece as `classical' can be misleading.  Moreover, in such examples, the part of the stress-energy tensor that remains nonzero in the $\hbar\rightarrow 0$ limit (in which the aforementioned cross terms of course vanish) is not the same as the classical stress energy tensor of the state's one-point function. 

The main issue that we will have to deal with in order to have an initial value formulation for semiclassical gravity concerns the definition of a suitable class of initial quantum states, $\omega_{\cal C}$ on the CCR algebra of the sharp-time field and field momentum (see Section \ref{Sect:star} for a definition) on an initial (would-be Cauchy) surface, $\cal C$.  Both in the context of quantum field theory in a fixed curved spacetime and also that of semiclassical gravity, it is usual to restrict attention to (globally hyperbolic) $C^\infty$ spacetimes
and to quantum states which satisfy the \emph{Hadamard condition} (see \cite[Chapter 3]{KW}).   The latter restriction is for at least two reasons:   First, states need to satisfy this condition, or rather a certain approximation to it (see e.g.\ \cite[Equations (102), (103)]{Decanini:2005eg}), for them to have a well-defined and finite expectation value for the renormalized stress-energy tensor.  Secondly \cite{FulSweWald, KW} in a general globally hyperbolic spacetime, the Hadamard property is conserved in the sense that states which are Hadamard in a neighbourhood of a Cauchy surface are necessarily Hadamard throughout the spacetime.  Thus one of our main aims here will be, for a given initial surface, $\cal C$, and a given quadruple, $(g_{ij}, \dot g_{ij}, g^{(2)}_{ij}, g^{(3)}_{ij})$, of classical gravitational initial data on that surface, to give a suitable \emph{surface Hadamard} definition such that, if an initial state, $\omega_{\cal C}$, exists, which satisfies it,
it would seem reasonable to conjecture that the Cauchy development $(({\cal M}, g_{ab}), \omega)$ (consisting of a spacetime, $({\cal M}, g_{ab})$ and a bulk Hadamard state $\omega$ which agree with the initial data on $\cal C$) will exist and be unique.

The difficulty is that the four-dimensional notion of `Hadamard' at a point depends on the geometry in a neighbourhood of that point, and, in particular, for a point in our initial surface, it will depend on all the time derivatives of $g_{ij}$ restricted to our surface, while the initial data only include $g_{ij}$ itself and its first three time derivatives.  It seems that this difficulty can however be overcome by using the semiclassical equations \eqref{semi-simple} and their successive time-derivatives (together with the equation \eqref{KG}) \emph{on} the initial surface so as to infer the missing time-derivatives.   

To illustrate how this works in a simpler context, we will first study, in Section \ref{Sect:YukSurf}, the analogous problem for our scalar Yukawa model.  From the equation, \eqref{phieq}, for the classical $\chi$ field (which expresses $\ddot \chi$ in terms of $\chi $ and its spatial derivatives)  one can take successive time derivatives and express  $\chi^{(3)}$ and higher derivatives  in terms of   spatio-temporal derivatives which involve lower time derivatives.  Using this approach we have shown for small $n$ and conjecture for all $n$ that the demand that the $n$th time derivative of $\chi$ (for $n >1$) be well-defined and finite on the initial surface, entails a condition on the initial state, $\omega_{\cal C}$ (namely that the appropriate Hadamard renormalization of the appropriate derivative of $\omega(\phi^2)$ exists\footnote{\label{ftnt:smooth} An important remark to be made here is that, on a $C^\infty$ background (so the coefficients, $V_n$, in the Hadamard series $\sum_nV_n\Sigma^n$ have well-defined finite values for all $n$)  should a state, $\omega$, of the $\phi$ field have a two-point function that failed to be exactly Hadamard, then, for some $n$, we would expect this procedure to fail and the $n$th time-derivative of $\chi$ not to exist.   We may conclude from this that in any formulation of the semiclassical scalar Yukawa model in which the states are only required to be approximately Hadamard, we would inevitably have to deal with solutions in which $\chi$ is not smooth.  And by a similar argument (cf.\ Note (B) after Equation \eqref{DecFolTay} in Section \ref{Sect:star})  any formulation of semiclassical gravity that admits states, $\omega$, which are not exactly Hadamard would involve nonsmooth metrics.}) which involves $\chi$ and time derivatives of $\chi$ restricted to the surface whose order is less than $n$.   Thus we arrive at a definition of a surface Hadamard initial state, as an initial state, $\omega_{\cal C}$, that both satisfies the resulting iteratively generated infinite sequence of conditions and which is also Hadamard in an obvious sense for the resulting iteratively generated sequence of values of $\chi$ and its successive time derivatives.
 
We will also study, in Section \ref{Sect:ElecSurf}, the analogous problems for semiclassical electrodynamics which is intermediate in difficulty between the scalar Yukawa model and semiclassical gravity. 

In generalizing the notion of surface Hadamard to semiclassical gravity, one encounters further difficulties.   First, for a pair of nearby points in our initial surface, the (squared) geodesic distance, $\Sigma$, which occurs in the Hadamard form (see Section \ref{Sect:star}) refers, in general, to geodesics that leave the surface and thus cannot immediately be inferred from the initial data.   However this difficulty may be resolved since for each of our needed successive approximations to the Hadamard form, it will suffice to expand $\Sigma$ in a covariant Taylor series to a suitable finite order in the (squared) geodesic distance \emph{in} the surface.   Secondly, when one mimics the iterative procedure of the scalar Yukawa model, it seems that the formula one obtains for the $n$th time derivative of the metric, $g_{ij}$, on the initial surface depends on the $n$th time derivative of the metric!  (As we will  see in Section \ref{Sect:ElecSurf} an analogous situation already occurs in semiclassical electrodynamics.)  In particular, for semiclassical gravity, the formula for the 4th time derivative of the metric (i.e.\ the first one that does not form part of the initial data) is given by consideration of the undifferentiated semiclassical Einstein equations -- where, as we shall explain in Section \ref{Sect:star}, the `higher derivative terms' $I_{ab}$ and $J_{ab}$ in the precise version \eqref{semi-hbar} of the right-hand side of Equation \eqref{semi-simple} include the terms $\Box R g_{ab}$ and $\Box R_{ab}$ which involve the fourth time-derivative of the metric -- and it is well-known (see again \cite[Equations (102), (103)]{Decanini:2005eg}) that, in order to obtain the right value for the expectation value of the term $\omega(T_{ab}^{(\ell){\rm ren,0}})$ in \eqref{semi-hbar}, one needs to retain terms in the Hadamard subtraction piece (in particular the covariant Taylor expansion of the Hadamard `$V$' coefficient) up to second order in the differentiated squared geodesic distance, $\Sigma_{,a}$, which \emph{also} involve $\Box R g_{ab}$ and $\Box R_{ab}$ and thus \emph{also} involve the 4th time derivative of the metric!  Thus $g^{(4)}_{ij}$ cannot easily be made the subject of \eqref{semi-simple}.   But, nevertheless, we shall argue that the semiclassical Einstein equations will still determine the 4th time derivative of the metric \emph{implicitly}.  And it seems that things go similarly for the $4+k$th time derivative of the metric  for which one needs to consider $k$ time derivatives of the semiclassical Einstein equations.   Thus, in the end, it seems that one may arrive at a definition for `surface Hadamard' just as we did for the scalar Yukawa model.   And we will conjecture that given initial data $g_{ij}$, $\dot g_{ij}$, $g^{(2)}_{ij}$, $g^{(3)}_{ij}$ for the classical gravitational field on an initial surface, $\cal C$, and an initial state $\omega_{\cal C}$ that satisfies the surface Hadamard condition with respect to those classical data, then there will exist a unique spacetime $({\cal M}, g_{ab})$ and a Hadamard state, $\omega$ (appropriate to the equation \eqref{KG}) on $({\cal M}, g_{ab})$ such that $\cal C$ is a Cauchy surface for $({\cal M}, g_{ab})$ and the restriction of $g_{ab}$ and its first 3 time-derivatives to $\cal C$ coincide with the initial gravitational data and $\omega$ restricts (in a sense we will make clear in Sections \ref{Sect:star} and \ref{Sect:SSC}) to $\omega_{\cal C}$ and the semiclassical Einstein equations, \eqref{semi-simple} hold (with right hand side spelled out as in \eqref{semi-hbar}).

It is important to note that the imposition of the full set of semiclassical equations on the initial surface does not only implicitly determine the 4th derivative of the metric but also entails that, if an initial state, $\omega_{\cal C}$, is surface Hadamard for some given quadruple of classical gravitational data, $(g_{ij}, \dot g_{ij}, g^{(2)}_{ij}, g^{(3)}_{ij})$, then, in particular, the constraint equations (i.e.\ the $0a$ components of \eqref{semi-simple} expressed in terms of $(g_{ij}, \dot g_{ij}, g^{(2)}_{ij}, g^{(3)}_{ij})$ and $\omega_C$) will be satisfied. Thus, in the case of semiclassical gravity, \emph{the surface Hadamard condition subsumes the constraint equations}.   (As we shall see in Section \ref{Sect:ElecSurf}, a similar statement is true for a similar reason for the Gauss Law constraint in semiclassical electrodynamics.)  In view of this (and just as in classical General Relativity, by no means all initial data will satisfy the constraint equations) we expect that the surface Hadamard condition will only be satisfiable for certain quadruples of classical gravitational initial data, $(g_{ij}, \dot g_{ij}, g^{(2)}_{ij}, g^{(3)}_{ij})$ and, indeed, that for many such quadruples, there will be no surface Hadamard states at all.

It remains an open question to exhibit quadruples, $(g_{ij}, \dot g_{ij}, g^{(2)}_{ij}, g^{(3)}_{ij})$ of classical initial data for which surface Hadamard initial states, $\omega_{\cal C}$, exist and to explore the sizes of the sets of such $\omega_{\cal C}$ when they do exist.

We will also introduce the notion, appropriate to physical solutions, of a `physical surface Hadamard' initial state $\omega_{\cal C}$ which we define (see Definition \ref{Def:PhysSurf}) to depend only on $g_{ij}$ and $\dot g_{ij}$ while clarifying that conjecture with the further conjecture (Conjecture \ref{Conj:PhysSurf}) that the requirement that the solution will be physical will then determine the initial values $g^{(2)}_{ij}$ and $g^{(3)}_{ij}$  in terms of $g_{ij}$ and $\dot g_{ij}$.  This is in line with our earlier conjecture (Conjecture \ref{Conj:Phys}) mentioned above that, for physical solutions, one needs to specify only the two pieces of classical Cauchy data $g_{ij}$ and $\dot g_{ij}$ (equivalently $g_{ij}$ and $K_{ij}$).

The notion of `surface Hadamard' appears to be considerably simpler and questions such as those above appear to be considerably easier if we content ourselves with (physical) solutions to order $\hbar$ which are Hadamard to order $\hbar$ and they are even simpler and easier if we content ourselves with solutions to order $\hbar^0$ which, as we remarked above, still go beyond classical general relativity.  For, as we explain in Sections \ref{Sect:Dephbar} and \ref{Sect:1andhbar}, the semiclassical equations then acquire a mathematical form much closer to that of the classical Einstein equations and involve neither higher derivative kinetic terms nor any need to get involved with the complications of the Hadamard condition.

\subsection{Conceptual considerations of semiclassical gravity with state collapse}
\label{Background}

As we have mentioned above, in \cite{EppleyHannah} and \cite{Page} a number of arguments have been given against semiclassical gravity in general and, in particular, against theories which attempt to combine semiclassical gravity with quantum state collapses.  We should mention already that the conclusions of \cite{EppleyHannah, Page} have been disputed in the literature, see e.g.\ \cite{Albers:2008as, Carlip:2008zf, Kent:2018epv, Mattingly:2005nna}, although we stress that these counterarguments do not conclude that gravity should not be quantised, but rather that semiclassical gravity should not be so easily discarded.

Eppley and Hannah \cite{EppleyHannah} argue that in semiclassical gravity with state collapses one might, by means of suitable measurements, be able to violate the uncertaintly principle and could arrange to send signals faster than light. Carlip \cite{Carlip:2008zf} (and see references therein) concludes that these problems are not decisive or at least not fatal to the usefulness of the theory. Mattingly \cite{Mattingly:2005nna} has questioned the feasibility of conducting the Eppley-Hannah experiment in our universe. Kent \cite{Kent:2018epv}, on the other hand, questions the validity of the argument if the gravitational field interacts with the state of quantum matter. We should add that the faster-than-light signaling objection would seem, anyway, not to apply to collapse models, where state collapses are stochastic, hence uncontrollable and unpredictable  \cite{Myrvold:2002}. 

In addition, Page and Geilker \cite{Page} remarked that, during the course of a measurement, the covariant conservation equation of the stress-energy tensor would be violated, and concluded, in our view, somewhat prematurely that only the Everett (Many Worlds)  formulation (which is free of state collapse) is consistent with semiclassical gravity.  The argument given in \cite{Page} is that, on the Everett interpretation, semiclassical gravity will avoid the violation of covariant conservation if one  limits consideration to the complete wave function, describing simultaneously all the Many  Worlds branches that the Everett interpretation  postulates.  But, as argued in \cite{Maudlin:2019}, the  problem reappears if one focuses, as is required in any concrete application, on any one of the individual  branches,  particularly the branch that we experience. 

Furthermore, Page and Geilker \cite{Page} also argue that  semiclasssical gravity without state collapses leads to results in contradiction with their non-null result for their  Schr\"odinger-cat-like Cavendish experiment.   The conclusion often drawn (and drawn in \cite{Page}) from that experimental result is that semiclassical gravity does not apply in such circumstances and that, to account for such situations, one needs to resort to full quantum gravity as well as the Everett formulation.

However, we expect that a theory that combines semiclassical gravity \emph{with state collapses} will turn out to be consistent with the non-null result for the Cavendish experiment.
\footnote{Furthermore it is worth emphazising that a completely satisfactory and truly working theory of quantum gravity is at the current time lacking in any event.}

It is true that, as mentioned in \cite{Page}, combining semiclassical gravity with collapses will entail that the covariant conservation of the expectation value of the quantum stress energy tensor is violated on collapse surfaces.   However here we do not regard this by any means as a fatal objection.   As noted before \cite{Maudlin:2019} some such violation is inevitable in any of the known paths to deal with the  measurement problem.   In fact, already in standard quantum mechanics, at moments of collapse, the unitary time-evolution rule is anyway violated, so we do not see the violation of covariant conservation as  truly surprising or objectionable. One can in fact adopt the view that, in analogy to, say, the Navier-Stokes description of fluids, the theory might be valid for some relatively long period of time, but interrupted at moments where  something takes place invalidating it for a short time, like, say, when an ocean wave breaks against the beach. 

At this stage, we should make note of an issue that often arises in treatments of quantum state reductions, be they induced by  measurements, or by a modified dynamics  involving spontaneous collapses, and which concerns the rules for the determination of  local observables or  more  appropriately, relevantly  and  generically,   the determination of {\it local beables}\cite{Bell-beables}. The issue  is that  if we are interested in  any  such  beable, taken to be  represented by say   a local operator $O(x)$ associated with a  spacetime  event $x$  (or a neighborhood thereof)  we must have a rule   to  extract a value for such a beable from a quantum  state  deemed to  characterize  the physical situation at hand.  The difficulty resides in the fact that, within those contexts,  the  quantum  states  are  associated  with  Cauchy surfaces (and depend on whether a surface is to the future or the past of a particular reduction event) and on the other hand,   through the  event $x$  pass  infinitely many  Cauchy hypersurfaces.  Thus, in general, the value assigned to the  local beable of interest  depends on the choice  of the surface passing through $x$ and  the corresponding  quantum state of the system  of interest (for instance  a quantum field). This issue  could have  significant implications  when considering semiclassical gravity  where the  expectation value of the stress-energy tensor is clearly being given a r\^ole of a local  beable (determining for instance  aspects of the  spacetime geometry at or around  $x$).  The issue has been considered extensively in works such as \cite{Albert-2015,Myrvold:2002} and there is  also an  interesting proposal to address it in \cite{Ghirardi-Grassi}.  (For  a related  idea see  \cite{Durr:2013asa}).

In the present work, however, these issues will not arise since we will work on the assumption that collapses occur on certain non-overlapping collapse surfaces, separating `epochs' in which no collapse occurs. So, while we assume that these collapse surfaces will occur at random according to some yet-to-be-formulated theory, for any single semiclassical solution with collapses there will never be any ambiguity as to which epoch any given event belongs.

We also hope that  the lessons learned from the present study will serve as a basis for developing methods capable of dealing with more complex collapse  theories which may fail to generate such a clean separation into epochs as we have just described.

What we do need to do, though, is to supply a rule for how the (derivatives of) the metric jump(s) on collapse surfaces and, as we discussed above, we shall start to address this in the present paper.

\subsection{Programme of the paper}
\label{subsec:Programme}

In Section \ref{Sect:star} we review the relevant elements of quantum field theory in curved spacetime and the theory of the renormalisation of the stress-energy tensor.  Much of this is well-known.  However some aspects of our discussion of states with nonvanishing one-point functions appears to be new, in particular, our discussion, in Section \ref{Sect:Dephbar}, of how the pieces of the Hadamard form depend on $\hbar$, as illustrated by our example, towards the end of that section, involving superpositions of coherent states.  We shall make use of the algebraic approach to quantum field theory, as it is well-adapted to curved spacetimes.  In Section \ref{Sect:SSC} we begin the discussion of  semiclassical gravity without quantum state collapses. Here, we introduce the notion of \emph{physical solutions} to the theory, defined to be one-parameter families of solutions, labelled by $\hbar$, which are smooth functions of $\hbar$ (jointly with coordinates) in a neighborhood of $\hbar=0$.   

A large part of Section \ref{Sect:SSC} is devoted to the definition of `surface Hadamard', which we first look at, in some detail, in Section \ref{Sect:YukSurf}, for our scalar Yukawa model.   We then point out that there are two technical obstacles that arise in attempting to generalize the notion of surface Hadamard from that model to semiclassical gravity. Full details on the technical issues that arise here will be given in the companion paper \cite{companion}. Others can be already illustrated in the context of semiclassical electrodynamics as we will discuss in Section \ref{Sect:ElecSurf}.

We formulate a number of conjectures about surface Hadamard states and then state conjectures about the well-posedness of the initial value problem for all three models.   In the case of semiclassial gravity, there is a conjecture (Conjecture \ref{Conj:GravityIV}) for general solutions, in which the classical initial data that needs specifying is assumed to consist of the restriction of the metric to an initial surface together with the restrictions of its first three time derivatives and a conjecture for physical solutions comprised of Conjectures \ref{Conj:Phys} and \ref{Conj:PhysSurf}) which posits that just the restrictions of the metric and its first time derivative (equivalently, the metric and the extrinsic curvature) need to be specified since, as we already remarked in the introduction, we expect that, for physical solutions, the second and third time derivative of the metric will be determined once the metric and its first time derivative are specified.   In Section \ref{Sect:1andhbar} we discuss the considerable simplifications in our conjectures that we seem to be able to make if, instead of exact solutions, we content ourselves with approximate solutions to our semiclassical equations to order $\hbar$, and with the even more drastic approximation of solutions to order $\hbar^0$.

Section \ref{Sect:collapse} then addresses the question of how the initial data for the classical gravitational field on a given collapse Cauchy surface needs to jump when some given abrupt change occurs to the quantum state on that surface.   This section begins by pointing out that, in the scalar Yukawa model, due of the absence of initial constraints, one may take the initial data to be continuous across the collapse surface, but that, for semiclassical electrodynamics,  the classical electric field must be discontinuous due to the sudden change, at a moment of quantum state collapse, in the expectation value of the electric charge density which appears in the Gauss constraint.   For the semiclassical Einstein equations, we arrive, by considering the analogy with the electromagnetic case, at the tentative proposal that the parts of the initial data $(g_{ij}, K_{ij})$ for a physical solution that do \emph{not} jump (i.e.\ are continuous also at collapse surfaces) are the full 3-metric $g_{ij}$ and the transverse traceless part (which we call $L_{ij}$) of the extrinsic curvature.  And we argue that the jump in the remaining part of the extrinsic curvature will then be determined by the change in the quantum state on collapse.

We shall adopt the same notational conventions as in \cite{Wald}, except that we do not set Newton's constant, $G_N$, to 1 and we use the letters $a, b, \ldots $ for (abstract) spacetime indices while $i, j, \ldots \ $ will indicate spatial indices. We use the notation $\dot{A}$ to indicate `time' derivative of $A$ in an appropriate sense. In particular, for initial data on a Cauchy surface, $\dot A$ indicates a derivative `off' the surface. We use the double-dot notation, $\ddot{A}$ in an analogous way. For $n$-th order `time' derivatives with $n \geq 3$ we use the notation $A^{(n)}$. In Gaussian normal coordinates the former notation makes reference to the derivative with respect to the time coordinate. $\nabla_a A$ denotes the covariant derivative of $A$ compatible with the spacetime metric $g_{ab}$. We also use the notation $A_{;a} = \nabla_a A$. $D_i B$ denotes the covariant derivative of $B$ compatible with the induced $3$-metric $g_{ij}$ on a spacetime surface, except in Appendix \ref{app:Y}, where we see $(M, h_{ij})$ as a $3$-dimensional Riemannian manifold in its own right and we continue to use the notation $\nabla_i A$ for the covariant differentiation of $A$. The differential operator $D_a$ (note the subindex $a$ as opposed to a middle-alphabet $i$) denotes a covariant derivative with Maxwell potential $A_a$. The symbol $\nabla$ without indices indicates the standard gradient operator and bold-face characters indicate $3$-dimensional vectors, e.g.\ ${\bf v}$. (This notation is used extensively when discussing electrodynamics.) Thus $\nabla.{\bf v}$ and $\nabla\times{\bf v}$ denote the divergence and curl of ${\bf v}$ respectively.

\section{Quantum field theory on a fixed curved spacetime and the expectation value of the renormalized stress-energy tensor}
\label{Sect:star}

We shall seek solutions to the semiclassical Einstein equations \eqref{semi-simple} (without collapses) in which the spacetime $({\cal M}, g_{ab})$ is globally hyperbolic and time-orientable.   In this section, we shall review the relevant facts from quantum field theory on a fixed such spacetime.   We take the viewpoint that a quantum field theory on such a spacetime is a unital *-algebra $\mathscr{A}({\cal M}, g_{ab})$, which associates algebra elements to spacetime regions in a local and covariant way, as first advocated by Dimock \cite{Dimock}. We adopt this prescription, as it is well adapted to general, curved spacetimes. Einstein causality is also assumed, whereby spatially separated observables commute.   Another desirable property is that if ${\cal C}$ is a Cauchy surface of ${\cal M}$, then for $\cal N$ a neighbourhood of ${\cal C}$, the subalgebra of $\mathscr{A}({\cal M}, g_{ab})$ generated by algebra elements associated to the region $\cal N$ is equal to the full algebra, $\mathscr{A}({\cal M}, g_{ab})$. This is known as the \emph{time-slice axiom}, and essentially means that the theory has a well-posed initial value problem.  Modern accounts of algebraic QFT and in particular of the so-called \emph{locally covariant QFT} can be found in \cite{Brunetti:2001dx,FV,FewRej}.

Quantum states are defined in the algebraic sense as maps $\omega: \mathscr{A}({\cal M}, g_{ab}) \to \mathbb{C}$, that satisfy (i) normalisation, $\omega(1\!\!1) = 1$, where $1\!\!1$ is the algebra unit; linearity, i.e.\ $\omega(c_1a_1 + c_2a_2) = c_1\omega(A_1) + c_1\omega(A_2)$;  and (ii) positivity, $\omega(a^* a) \geq 0$ for any $a_1, a_2, a \in \mathscr{A}({\cal M}, g_{ab})$.  The connection with the more familiar Hilbert space approach (see e.g.\ \cite{Haag, KW, FewRej}) is illustrated by the fact that, in a representation, $r$, of the *-algebra by operators on a Hilbert space, $\cal H$, examples of states are obtained as expectation values; in particular, the formulae $\omega(a) = \langle\Psi|r(a)\Psi\rangle$ for a choice of vector, $\Psi\in {\cal H}$ or
$\omega(a) = {\rm tr}(\rho r(a))$ for a density operator, $\rho$ on $\cal H$ both define states in the algebraic sense.  

In order to define the right hand side, $\omega(T_{ab})$, of \eqref{semi-simple}, we will need a notion of the expectation value of the stress-energy tensor.  If the quantum field theory arises from the quantisation of a quadratic Lagrangian field theory, the stress-energy tensor of the quantum theory can be defined through a point-splitting renormalisation scheme based on suitable derivatives of products of unsmeared fields and with the appropriate subtraction of coincidence-limit singularities, with the additional imposition of covariant conservation.  The renormalized stress-energy tensor will not be an element of the *-algebra $\mathscr{A}({\cal M}, g_{ab})$, but one can enlarge the algebra suitably to include such non-linear observables into an interacting algebra (see Chapter 3 in \cite{KW} and also \cite{Hollands:2001nf, Hollands:2001fb, Hollands:2004yh, KhavMor}). Such enlarged algebras are also useful for dealing perturbatively with interacting theories \cite{Rejzner}. 

To make sense of Equation \eqref{semi-simple}, however, it suffices to be able to define the \emph{expectation value of the renormalized stress-energy tensor}, and it will not be necessary to have a definition of the renormalized stress-energy tensor itself as an element of any particular algebra.

Returning to our real Klein-Gordon field satisfying \eqref{KG}, we may, as is done by many authors (see e.g.\ \cite{KhavMor}), define the unital *-algebra, $\mathscr{A}({\cal M}, g_{ab})$ to be generated by smeared fields $\phi(F)$ with $F\in C_0^\infty({\cal M})$, formally $\int_M \! \phi(x) F(x)\, \sqrt{g}\,d^4x$, which satisfy the equivalence relations
\begin{enumerate}[(i)]
\item linearity, $F \mapsto \phi(F)$ is linear;
\item hermiticity, $\phi(F)^* = \phi(F)$;
\item field equation, $\phi(G) = 0$, if $G = (\Box - m^2 - \xi R) F$;
\item commutation relations, $[\phi(F), \phi(H)] = -i E(F,H)1\!\!1$, for $H \in C_0^\infty({\cal M})$,
\end{enumerate}
where $E = E^- - E^+$ is the advanced-minus-retarded fundamental solution of $(\Box - m^2 - \xi R)$. The Klein-Gordon algebra, $\mathscr{A}({\cal M}, g_{ab})$, is the unital *-algebra generated by the equivalence classes of smeared fields $\phi(F)$ subject to relations (i)-(iv). The generators can be identified locally and covariantly under metric embeddings by pushforward.  (See \cite{FV} for details.) The time-slice axiom follows from the well-posedness of the initial value problem for the Klein-Gordon equation, and in fact more is true, namely the existence of sharp-time fields, $\varphi=\phi|_{\cal C}$ and $\pi = \nabla_n\phi|_{\cal C}$ on each Cauchy surface ${\cal C}$.  Here, $\nabla_n$ denotes the derivative with respect to the future-pointing unit normal $n^a$ to ${\cal C}$.  
Following the standard treatments, we introduce the symplectic product  on elements  of  phase  space  represented  by  pairs $(f,p)$  through $\varsigma(f_1, p_1; f_2, p_2) = \int ( f_1p_2-p_1f_2 ) {\sqrt{\det ^3\!g}}\,d^3x$ where  $^3\!g$ is  the determinant of the induced 3-metric, $g_{ij}$, on ${\cal C}$.   The  sharp-time fields $\varphi, \pi$  will satisfy the canonical commutation relations (\emph{CCR}) which may be expressed neatly in terms of \emph{symplectically smeared} Cauchy data (see e.g.\ \cite{KW}) $\varsigma(\varphi, \pi; f, p) = \varphi(p) -\pi(f)$,
where for a classical solution, $\phi_{\rm class}$, $f = \phi_{\rm class}|_{\cal C}$ and $p = \nabla_n \phi_{\rm class}|_{\cal C}$. The CCR are then  expressed  by: 
\begin{equation}
\label{CCR}
[\varsigma(\varphi, \pi; f_1, p_1), \varsigma(\varphi, \pi; f_2, p_2)] = i\varsigma(f_1, p_1; f_2, p_2)1\!\!1.
\end{equation}
 
It will be important for our treatment of the initial value problem for semiclassical gravity in Section \ref{Sect:SSC} that the CCR *-algebra, which we shall call $\mathscr{A}({\cal C}, ^3\!g)$, generated by the $\varsigma(\varphi, \pi; f, p)$ satisfying \eqref{CCR} is isomorphic to the *-algebra $\mathscr{A}({\cal M}, g_{ab})$ by the isomorphism ${\rm iso}_{\cal C}$ defined so that 
\begin{equation}
\label{iso}
{\rm iso}_{\cal C}: \varsigma(\varphi, \pi; f, p) \mapsto \phi(F)
\end{equation}
where $(f, p)$ are the Cauchy data on ${\cal C}$ of the classical solution $\phi_{\rm class} := EF$.  (This is well-defined because, if $EF_1=EF_2$, then $F=F_1-F_2$ takes the form  $(\Box - m^2 - \xi R)G$ for some $G\in C_0^\infty({\cal M})$ (namely for $G=E^-F=E^+F$) and hence, by (iii) above, $\phi(F_1)=\phi(F_2)$.)  (We write $\mathscr{A}({\cal C}, ^3\!g)$, to indicate that our definition of $\mathscr{A}({\cal C}, ^3\!g)$ depends on the determinant, $^3\!g$ of the 3-metric, $^3\!g_{ij}$ induced on ${\cal C}$ by $g_{ab}$ since the square root of that determinant appears in the volume element in the definition of $\varsigma$.)

Let us note here that specifying the symplectically smeared two-point function in a state, $\omega_{\cal C}$, on the CCR algebra of a Cauchy surface, $\cal C$ (i.e.\ the expectation value of the product of a pair of symplectically smeared fields) is tantamount to specifying smeared versions of the four quantities
\begin{equation}
\label{four2points}
\omega_{\cal C}(\varphi(x)\varphi(x')), \ \omega_{\cal C}(\varphi(x)\pi(x')), \ \omega_{\cal C}(\pi(x)\varphi(x')), \ \omega_{\cal C}(\pi(x)\pi(x')).
\end{equation}
for all pairs, $x, x'\in {\cal C}$.  And let us also note that, as long as $x$ and $x'$ are distinct, $\varphi(x)$ and $\pi(x')$ will commute and therefore
$\omega(\varphi(x)\pi(x')) = \omega(\pi(x')\varphi(x))$ and thus the third item in the above list is reduntant.  (And this is still true in general since the difference will be $\delta^{(3)}(x,x')$.)  

Einstein causality is a direct consequence of (iv). 

We will need to restrict the set of all states, $\omega: \mathscr{A}({\cal M}, g_{ab}) \to \mathbb{C}$, to a subset which have a suitable short-distance behavior to permit the definition of a renormalized stress-energy tensor.   In the context of quantum field theory in a fixed curved spacetime, this subset is usually taken to be the set of \emph{Hadamard states} and we will recall the definition next. (For more about the definition of Hadamard states, see \cite{KW} and the recent comment on that \cite{NewMoretti}.   See also \cite{KhavMor}.   Let us also mention that there is an important microlocal reformulation of the Hadamard condition, introduced in \cite{Radz}.  However we will not make use of that in the present work.)

A state is said to satisfy the \emph{Hadamard condition} if, in any convex normal neighbourhood, ${\cal N} \subset {\cal M}$, the (integral kernel of the) Wightman two-point function has the form:
\begin{align}
\omega(\phi(x) \phi(x')) = \frac{1}{8\pi^2} \left( \frac{U(x, x')}{\Sigma} +  V(x, x') \ln \left(\frac{\Sigma(x, x')}{\ell^2} \right) + W_{{\ell} }^{(\omega)}(x, x') \right)
\label{HadamardCondition}
\end{align}
where  $U = \Delta^{1/2}$ and $\Delta$ is the van Vleck-Morette determinant given by
\begin{equation}
\label{vanVleck}
\Delta(x, x') = - (-g(x))^{-1/2}{\rm det}(-\Sigma_{;ab})(-g(x'))^{-1/2},
\end{equation}
${\ell} \in \mathbb{R}^+$ is a length scale, $\Sigma(x, x')$ is half the squared geodesic distance between spacetime points $x, x' \subset N$ and  $V$ and $W_{{\ell}}^{(\omega)}$ are symmetric, smooth bi-functions, 
formally defined by the Hadamard recursion relations, which are obtained (see e.g.\ \cite{DeWittBrehme,Decanini:2005eg}) by writing out the formal expansions
\begin{subequations}
\label{vwHad}
\begin{align}
V(x, x') & = \sum_{n = 0}^\infty V_n(x, x') \Sigma^n(x, x'), \label{vcoeff}\\
W_\ell^{(\omega)}(x, x') & = \sum_{n = 0}^\infty W_{\ell \, n}^{\omega}(x, x') \Sigma^n(x, x'), \label{wcoeff}
\end{align}
\end{subequations}
and using the fact $\omega(\phi(x) \phi(x'))$ solves the Klein-Gordon equation in $x$. In Equation \eqref{wcoeff} $W_{\ell 0}^\omega$ is not fixed by the Hadamard recursion relations; instead it is fixed by the state's two-point function, and indeed characterises that two-point function.  On the other hand, $V(x, x')$, like $U(x, x')$, is independent of the state.

We note that, given a fixed Hadamard state (unless $V(x, x')$ vanishes) there is an ambiguity in the decomposition of the two-point function into singular and regular pieces (cf.\ Equation \eqref{HadamardCondition}), since a change in the choice of ${\ell}$ to, say $\ell'$, can be compensated by replacing $W_{{\ell} }^{(\omega)}$ by $W_{{\ell'} }^{(\omega)} = W_{{\ell} }^{(\omega)} + V\ln(\ell'^2/\ell^2)$, which is the reason why we include the $\ell$-label in $W_{{\ell} }^{(\omega)}$ -- often omitted in the literature.

In order to obtain the  renormalized  expectation  value of the stress-energy  tensor in any Hadamard  state,  we introduce  the \emph{Hadamard subtraction piece}, $H_\ell$:
\begin{align}
H_\ell(x, x') := \frac{1}{8\pi^2} \left( \frac{U(x, x')}{\Sigma} +  V(x, x') \ln \left(\frac{\Sigma(x, x')}{\ell^2} \right) \right)
\label{Hadamard}
\end{align}
where $U$, $\ell$, $\Sigma$ and $v$ are as above. ($H_\ell$ corresponds to what is called in \cite{Decanini:2005eg}, whose account we are broadly following, the `singular part' of the two-point function, but note that the discussion there is based on the Feynman two-point function instead of our Wightman function and most of the formulae there suppress the dependence on the length scale, called here $\ell$.) 

The expectation value of the stress-energy tensor in a Hadamard state\footnote{\label{twopoint} We remark that, in view of this formula, the expectation value of the stress-energy tensor depends only on the two-point function of the state $\omega$.}, $\omega$, is given by the formula \cite{Wald78,Decanini:2005eg} 
\begin{equation}
\label{Tabren}
\omega(T_{ab}^{(\ell){\rm ren}}(x) ) = \lim_{x' \to x} \mathscr{T}_{ab}(x, x') \left[\omega( \phi(x) \phi(x') )- H_\ell(x, x') \right] + \frac{1}{4 \pi^2} g_{ab} [V_1](x) + \Theta_{ab}(x),
\end{equation}
and we will explain next how the various terms, other than $H_\ell$, in this formula are defined and why they are there.  

Before doing so, let us remark that, as explained in \cite{Decanini:2005eg}, as far as its role in \eqref{Tabren} is concerned, we may as well replace $H_\ell$ by the approximate form, which we shall call here $H^{\rm DF}_\ell$, in which the Hadamard expansion, \eqref{vcoeff}, of the $V$ term is replaced by its first two terms, $V_0 + V_1\Sigma$ and also the covariant Taylor expansions\footnote{\label{ftnt:covTay} We adopt the convention in covariant Taylor expansions that the coefficients are functions of $x$.  Thus e.g.\ if we indicate the arguments, the first equation of \eqref{DecFolTay} will begin $U_0(x,x') = u_0(x) + u_{a}(x)\Sigma^{;a}(x,x') + \dots$. Also the covariant derivatives of $\Sigma$ are taken with respect to its first argument. Let us also note that one may easily consolidate the covariant Taylor expansions of $V_0$ and $V_1$ into a single covariant Taylor expansion of $V= V_0 + V_1\Sigma$ by using the identity $\Sigma = \frac{1}{2} g_{ab}\Sigma^{;a}\Sigma^{;b}$.}
\begin{subequations}
\label{DecFolTay}
\begin{align}
U &= u_0 + u_{a}\Sigma^{;a} + \frac{1}{2!}u_{ab}\Sigma^{;a}\Sigma^{;b} +  \frac{1}{3!}u_{abc} \Sigma^{;a}\Sigma^{;b} \Sigma^{;c} + \frac{1}{4!}u_{abcd}\Sigma^{;a}\Sigma^{;b} \Sigma^{;c}\Sigma^{;d} + O(\Sigma^{5/2}),\label{UcovTay}\\ 
V_0 &= v_0 + v_{0a}\Sigma^{;a} + \frac{1}{2!}v_{0ab}\Sigma^{;a}\Sigma^{;b} + O(\Sigma^{3/2}),\label{V0covTay}\\ 
V_1 &= v_{10} + O(\Sigma^{1/2}) \label{V1covTay}
\end{align}
\end{subequations}
of $U$, $V_0$ and $V_1$ are truncated at their second order, first order and zeroth order terms in $\Sigma$ respectively.  (See \cite[Equations (104), (105), (106)]{Decanini:2005eg} but note that our sign convention for the coefficients differs from that in \cite{Decanini:2005eg}, where the signs alternate.)   This is because the contribution of the omitted terms in these expansions to $H_\ell$ will tend to zero in the limit in \eqref{Tabren}.  Note however (and both points (A) and (B) which follow will be relevant to our definition of `surface Hadamard' in Section \ref{Sect:SSC}): 

\medskip

\noindent
(A) that the second order terms in the expansion of $V_0$ and the zeroth order terms in the expansion of $V_1$ contain (see \cite[Equations (108c), (109))]{Decanini:2005eg}) terms, namely $\Box R$ and $\Box R_{ab}$ that, expressed in terms of Gaussian normal coordinates for an initial surface, $\cal C$, involve \emph{4th} time-derivatives of $g_{ij}$;

\medskip

\noindent
(B) that were we to calculate successively higher time derivatives of $\omega(T_{ab}^{(\ell){\rm ren}}(x) )$, by differentiating the right hand side of \eqref{Tabren}, then we would need to include successively more terms in the Hadamard expansion \eqref{vcoeff} and successively more terms in the covariant Taylor expansions, \eqref{DecFolTay}, for $U$ and $V_0$, $V_1$, $\dots$ which will, in turn, involve ever higher time derivatives of $g_{ij}$.

\medskip

$\mathscr{T}_{ab}$ in \eqref{Tabren} is the bidifferential operator appropriate to the point-split form of the classical stress-energy tensor for the (nonminimally coupled) Klein-Gordon field, say $\chi$. The classical stress energy tensor is
\begin{align}
T_{ab} = (1 -2 \xi) \chi_{;a} \chi_{;b} + \left(2 \xi - \frac{1}{2}\right) g_{ab} g^{cd} \chi_{;c} \chi_{;d} - 2 \xi \chi \chi_{;ab} + 2 \xi g_{ab} \chi \Box \chi  + \xi G_{ab} \chi^2 - \frac{1}{2} g_{ab} m^2 \chi^2.
\label{TabClass}
\end{align}

Defining $\mathscr{T}_{ab}$ as
\begin{align}
\mathscr{T}_{ab}(x, x') &  := (1-2\xi ) g_{a}\,^{b'} \nabla_a \nabla_{b'} +\left(2\xi - \frac{1}{2}\right) g_{ab}g^{cd'} \nabla_c \nabla_{d'}  - \frac{1}{2} g_{ab} m^2  \nonumber \\
& + 2\xi \Big[  - g_{a}\,^{a'} g_{b}\,^{b'} \nabla_{a'} \nabla_{b'} + g_{ab} g^{c d}\nabla_c \nabla_d + \frac{1}{2}G_{ab} \Big],
\label{TabPointSplit}
\end{align}
where $g_a\,^{b'}$ is the bi-vector of parallel transport, which parallel transports vectors at $x'$ to the point $x$ (see \cite{DeWittBrehme,Decanini:2005eg}), one can guarantee that applying the operator \eqref{TabPointSplit} to the point-split product of classical fields $\chi(x)\chi(x')$ and taking the limit $x' \to x$ yields the classical stress-energy tensor, Equation \eqref{TabClass}. 

Note that there are several possible reasonable choices one can make for the point-splitting operator 
$\mathscr{T}_{ab}(x, x')$ \eqref{TabPointSplit}.  Obviously there are many possible definitions that, at a classical level with converge to the classical stress-energy tensor in the coincidence limit.   But also, given that, in \eqref{Tabren}, $\mathscr{T}_{ab}(x, x')$ acts on  $\omega(\phi(x) \phi(x')) - H_\ell(x,x')$ which is both smooth and symmetric under interchange of $x$ and $x'$, there is also considerable leeway to modify $\mathscr{T}_{ab}(x, x')$ without changing the result for the limit in \eqref{Tabren}.   For example, one may easily see that this limit would be unchanged if, in a given coordinate chart, we were to replace the bitensor $g_a\,^{b'}$ by a simple delta function $\delta_a\,^{b'}$.   Moreover, one could, e.g.\  replace the term  $- g_{a}\,^{a'} g_{b}\,^{b'} \nabla_{a'} \nabla_{b'}$ by  $-\nabla_{a} \nabla_{b}$.

$[V_1]$ in Equation \eqref{Tabren} denotes the coincidence limit of the coefficient $V_1$ in the expansion \eqref{vcoeff} ($=$ the $v_{10}$ of \eqref{DecFolTay}) and
the term that contains it, $\frac{1}{4 \pi^2} g_{ab} [V_1](\x)$ is needed (see \cite{Wald78,Decanini:2005eg}) to ensure covariant conservation.  It is also responsible for the trace anomaly of the renormalized stress-energy tensor of the conformally-coupled Klein-Gordon field, with $m^2 = 0$, $\xi = 1/6$.  Written out in terms of the local geometry, we have
\begin{align}
\frac{1}{4 \pi^2} g_{ab} [V_1](\x) & = \frac{1}{4\pi^2} g_{ab} \left( \frac{1}{8} m^4 + \frac{1}{4} \left(\xi - \frac{1}{6}\right) m^2 R - \frac{1}{24}\left(\xi - \frac{1}{5} \right) \Box R  \right. \nonumber \\
& \left. + \frac{1}{8} \left(\xi - \frac{1}{6} \right)^2 R^2 - \frac{1}{720} R_{ab} R^{ab} + \frac{1}{720} R_{abcd} R^{abcd} \right).
\label{TraceAnom}
\end{align}

Finally, $\Theta_{ab}$ is given by
\begin{subequations}
\label{Theta}
\begin{align}
\Theta_{ab} & :=  \alpha_1 g_{ab} + \alpha_2 G_{ab} + \alpha_3 I_{ab} + \alpha_4 J_{ab}, \\
I_{ab} & := \frac{1}{\sqrt{- g}}\frac{\delta}{\delta g_{ab}}\int R^2 \sqrt{-g}\,d^4x\, 
= 2R_{;ab} - 2 R R_{ab} + g_{ab}(-2\square R + \frac{1}{2}R^2), \\
J_{ab} & :=  \frac{1}{\sqrt{-g}}\frac{\delta}{\delta g_{ab}}\int R_{cd}R^{cd}{\sqrt{-g}}\,d^4x\, 
=R_{ab}-\square R_{ab} - 2 R^{cd}R_{acbd} + g_{ab}(-\frac{1}{2}\square
 R + \frac{1}{2}R_{ab}R^{ab}),
\end{align}
\end{subequations}
(where $\alpha_1, \ldots, \alpha_4 \in \mathbb{R}$ denote arbitrary constants, $\alpha_1$ having the  dimensions of the inverse fourth power of a length, $\alpha_2$ of the inverse squared power of a length, while $\alpha_4$ and $\alpha_4$ are dimensionless) is the most general symmetric, covariantly conserved, local, rank-$(0,2)$ tensor constructed out the metric and its derivatives, with dimension of inverse length to the fourth power, and expresses the ambiguity (see \cite{Wald78,Decanini:2005eg}) in the renormalized stress energy tensor.

We remark that one can already see the necessity of including at least a part of the term $\Theta_{ab}$ by noticing that (unless $V(x, x')$ vanishes) even if we were to set $\Theta_{ab}$ to zero for one choice of the arbitrary quantity (with dimensions of length) $\ell$ in \eqref{Hadamard}, then, a $\Theta_{ab}$ term would (re-)appear, with some particular linear combination of $\alpha_1,\dots, \alpha_4$ were we to change $\ell$ to some other value, say $\ell'$. 
The details of the expression are given by Equation (79) in \cite{Decanini:2005eg} setting $M^2 = (\ell/\ell')^2$ in that formula.

Also while regularisation and renormalisation schemes other than the point splitting approach suggest privileged values for these constants, different schemes suggest different values.   So none of the values is to be taken too seriously a priori as privileged.   Rather (in the absence of input from some theory of quantum gravity) and in line with general practice, we regard all choices of $\alpha_1, \dots, \alpha_4$ as on an equal footing.   In any case, and this is important to emphasize, whichever viewpoint we take, and whatever value $\ell$ is taken to have, and whatever values we choose for $\alpha_1 \dots \alpha_4$, for a fixed choice of $\ell$, the semiclassical Einstein equations with $\omega(T^{(\ell)\rm ren}_{ab})$ on the right hand side will always have the character of a 4th order equation in the sense that a general solution would, we expect, require the setting of both the induced metric and the first three derivatives of the metric on a Cauchy surface.  (See Section \ref{Sect:SSC}.)

When we couple the expectation value of the stress-energy tensor to gravity according to \eqref{semi-simple} the contributions proportional to the coefficients $\alpha_1$ and $\alpha_2$ can obviously be absorbed into a redefinition (i.e.\ finite renormalization) of the cosmological constant, $\Lambda$, and Newton's constant, $G_{\rm N}$ in \eqref{semi-simple}, and thus, without loss of generality, one may assume that $\alpha_1$ and $\alpha_2$ are zero and can then also obviously redefine $\alpha_3$ and $\alpha_4$ so that the new semiclassical Einstein equations continue to take the form \eqref{semi-simple}.   We shall assume that done from now on.    One could similarly set (the new) $\alpha_3$ and $\alpha_4$ to zero at the expense of including higher derivative terms proportional to $I_{ab}$ and $J_{ab}$ on the left hand side of the semiclassical Einstein equations, prefaced by two new coupling constants and by similarly absorbing $\alpha_3$ and $\alpha_4$ into redefinitions (i.e.\ finite renormalizations) of those constants.   However, we prefer here not to take that view, but rather we shall assume some fixed choice of $\ell$ has been made, and keep the terms proportional to $\alpha_3$ and $\alpha_4$ on the right hand side of the semiclassical Einstein equations.  We could take the view that, although the correct values of $\alpha_3$ and $\alpha_4$  are unknown to us,  they will one day be fixed (given a fixed choice of $\ell$ in \eqref{Hadamard}) by a yet-to-be formulated quantum gravity theory of which our semiclassical gravity (with state collapses) is an approximation.

Adopting that point of view, we may now give a clear statement of how the right hand side of the semiclassical Einstein equations, \eqref{semi-simple}, is to be defined.  Namely,
\begin{equation}
\label{semi-hbar}
\omega(T_{ab}^{\rm ren} ) := \omega(T_{ab}^{(\ell){\rm ren,0}}) +
\hat\alpha_3 I_{ab} + \hat\alpha_4 J_{ab},
\end{equation}
where $\omega(T_{ab}^{(\ell){\rm ren,0}})$ is given by \eqref{TabPointSplit} with $\Theta_{ab}=0$ and $\hat\alpha_3$ and $\hat\alpha_4$ are constants.

\subsection{\label{Sect:Dephbar} Dependence on $\hbar$}

In preparation for the discussion of `physical solutions' in Section \ref{Sect:SSC} and for the discussion about solutions to order $\hbar^0$ and solutions to order $\hbar$ in Section \ref{Sect:Dephbar}, we next wish to ask how $\omega(T_{ab}^{\rm ren}(x) )$  depends on $\hbar$.   It is clear on dimensional grounds that, when we no longer set $\hbar$ to one, then the terms $\frac{1}{4 \pi^2} g_{ab} [V_1](x)$ and $\Theta_{ab}(x)$ in \eqref{Tabren} acquire $\hbar$ prefactors as therefore also does the term $\hat\alpha_3 I_{ab} + \hat\alpha_4 J_{ab}$ in \eqref{semi-hbar}.   Here we need to note that, if one were to proceed naively, when one restores $\hbar$ to all our equations, one would also replace the mass term, $m^2$, in the Klein Gordon equation \eqref{KG} by $m^2/\hbar^2$ and one might think that the absorption of the $\alpha_1$ and $\alpha_2$ into the definitions of $\Lambda$ and $G_N$ as explained above might also lead to a dependence on $\hbar$ for these two physical constants.  However, we shall adopt the view here that neither the mass term in \eqref{KG} nor $\Lambda$ nor $G_N$ depend on $\hbar$.   This is in line with the traditional assumptions that are made when one says that the loop expansion in perturbation theory is an expansion in powers of $\hbar$ although it does have the strange consequence that the $m^2$ term in \eqref{KG} -- which we choose not to divide by $\hbar$ -- retains the dimensions of an inverse length squared rather than of mass squared even though, once we no longer set $\hbar$ equal to one (and continue not to set $G_N$ to one) these no longer have the same dimensionality.

To find out how the full $\omega(T_{ab}^{{\rm ren,0}})$ depends on $\hbar$, we need to ask how the terms in the Hadamard form \eqref{HadamardCondition} depend on $\hbar$.   It may be seen also on dimensional grounds that the singular terms acquire an $\hbar$ prefactor, but what about the smooth term, $W_{{\ell} }^{(\omega)}$?   If $\omega_0$ is a \emph{quasifree state} with vanishing one-point function (in the pure case, this is also known as a `frequency-splitting vacuum' -- for this and the definition of quasifree state in the context of the Klein Gordon equation in curved spacetimes, see e.g.\ \cite{KW, KhavMor} ) then it is well-known and easy to see that $W_{{\ell} }^{(\omega_0)}$ too will be proportional to $\hbar$.   (This is to be expected since one may then regard it as a `one-loop' term.)    On the other hand, for a quasifree state, $\omega_1$, with a nonvanishing one-point function, one finds that $W_{{\ell} }^{(\omega_1)}$, and in consequence $\omega_1(T^{\rm ren}_{ab})$, will be the sum of a `quantum piece' proportional to $\hbar$ and a `classical-like' piece which is independent of $\hbar$.  To see this, let us note that we may think of such an $\omega_1$ as arising from a quasi-free state, $\omega_0$, with vanishing one-point function by the formula $\omega_1 = \omega_0\circ \gamma$ where $\gamma$ is the automorphism generated by the mapping $\phi\mapsto \phi + \phi_{\rm class}1\!\!1$ for some classical solution, $\phi_{\rm class}$.  In this case, the one-point function, $\omega_1(\phi)$ is easily seen to equal the classical solution, $\phi_{\rm class}$, and we may easily see that the two-point function, $\omega_1(\phi(x)\phi(x'))$ is given by
\begin{equation}
\label{trunc}
\omega_1(\phi(x)\phi(x')) = \omega_0((\phi(x) + \phi_{\rm class}(x)1\!\!1)(\phi(x') + \phi_{\rm class}(x')1\!\!1))
= \omega_0(\phi(x)\phi(x')) + \phi_{\rm class}(x)\phi_{\rm class}(x')
\end{equation}
\eqref{trunc} shows that $\omega_0$ is the \emph{truncation} (see e.g.\ \cite{Haag}) of $\omega_1$.   We also notice that the difference between $\omega_1$ and $\omega_0$ is smooth and that $\omega_0$ is of order $\hbar$ while the classical smooth piece of the two-point function,  $\phi_{\rm class}(x)\phi_{\rm class}(x')$, does not depend on $\hbar$.

In consequence of \eqref{trunc}, it is easy to see from the definition \eqref{Tabren} that we will have
\[
\omega_1(T^{\rm ren}_{ab}) = \omega_0(T^{\rm ren}_{ab}) + T^{\rm class}_{ab}
\]
where $T^{\rm class}_{ab}$ denotes the classical stress-energy tensor of the classical solution 
$\phi_{\rm class}$.   So $\omega_1(T^{\rm ren}_{ab})$ is the sum of a piece proportional to $\hbar$ and a piece independent of $\hbar$.

However, and as far as we are aware this has not been pointed out hitherto, there are states, $\omega$, for which $W_{{\ell} }^{(\omega)}$, and in consequence, $\omega(T^{\rm ren}_{ab})$, have a more complicated dependence on $\hbar$.   To give a class of examples, let us first note that if a quasifree state, $\omega_1$, with nonvanishing one point function as above is such that the classical solution (i.e.\ the state's one-point function) $\phi_{\rm class}$ has (say) compact support on Cauchy surfaces, then $\omega_1$ will arise as a coherent state in the (GNS) Hilbert space representation, $r$, of the *-algebra $\mathscr{A}({\cal M}, g_{ab})$ associated to its truncation, $\omega_0$.  (For the GNS representation, see e.g.\ \cite{Haag, FewRej}; for details on quasifree states of the Klein Gordon equation in curved spacetimes, see \cite{KW} and especially Appendix A there.   See also \cite{KhavMor}.)  We will, indeed, have that, for any operator, $A$, in $\mathscr{A}({\cal M}, g_{ab})$,  $\omega_1(A) = \langle\Psi(\phi_{\rm class}) |A\Psi(\phi_{\rm class})\rangle$ where $\Psi(\phi_{\rm class})$ is the coherent state vector 
\[
\Psi(\phi_{\rm class}) = e^{\frac{ir(\varsigma(\phi, \phi_{\rm class}))}{\hbar}}\Omega
\]
in the representation Hilbert space, where $\Omega$ is the GNS vacuum vector for $\omega_0$ and $\varsigma(\phi, \phi_{\rm class})$ is, on any Cauchy surface, ${\cal C}$, equal to
$\varsigma(\varphi, \pi; f, p)$ where $(f,p)$ are the Cauchy data on that surface of the classical solution $\phi_{\rm class}$.

Now, for a fixed such $\omega_0$, we can consider the state, let us call it $\omega_{\rm super}$ defined by the expectation value, $\langle\Psi^{\rm super}|\cdot\Psi^{\rm super}\rangle$, in a superposition, $\Psi^{\rm super}$, of two such coherent state vectors, 
\[
\Psi^{\rm super}= N(c_1\Psi(\phi_{\rm class}^1) + c_2\Psi(\phi_{\rm class}^2))
\]
where $N$ is the appropriate normalization factor.   We remark that $N^{-2}$ is not simply $|c_1|^2 + |c_2|^2$ because $\Psi(\phi_{\rm class}^1)$ and $\Psi(\phi_{\rm class}^2)$ will not be orthogonal.  Clearly, the two point function in such a state will be the ratio of the unnormalized two-point function (i.e.\ $N^{-2}$ times the normalized two-point function) to the inverse square, $N^{-2}$, of the normalization factor.   Each of these quantities will include cross terms -- i.e.\ $c_1^*c_2\langle\Psi(\phi_{\rm class}^1)|\phi(x)\phi(x')\Psi(\phi_{\rm class}^2)\rangle$ (plus its complex conjugate) for the unnormalized two-point function and  $c_1^*c_2\langle\Psi(\phi_{\rm class}^1)|\Psi(\phi_{\rm class}^2)\rangle$ (plus its complex conjuagate) for $N^{-2}$  -- and it is easy to see that each of these cross terms will be of the form $A\exp(-B/\hbar)$ where $A$ and $B$ are real-valued expressions independent of $\hbar$ with $B$ positive.   The result will, nevertheless, be that, at least within this class of examples, while the two-point function, and hence also $W_{{\ell} }^{(\omega_{\rm super})}$, and hence also $\omega_{\rm super}(T^{\rm ren}_{ab})$, will not be simply a sum of a piece independent of $\hbar$ and a piece proportional to $\hbar$, it (and hence also the stress-energy tensor) will still be the sum of a piece of order $\hbar^0$ and a piece of order $\hbar$.  Also a short calculation shows that the piece of the two-point function which is of order $\hbar^0$ will be smooth and the piece of order $\hbar$ will have the usual Hadamard singularity.  (Note though that, in this example, and unlike the case, discussed above, of quasi-free states with nonvanishing one-point function, the part of the stress-energy tensor that remains nonzero in the $\hbar\rightarrow 0$ limit is not the same as the classical stress energy tensor of the state's one-point function.)  

Partly motivated by the above class of examples, when we discuss \textit{physical solutions} to semiclassical gravity in Section \ref{Sect:SSC}, the most general class of states, $\omega$, that we shall consider will be those for which the two-point function $\omega(\phi(x)\phi(x'))$ takes the form $G_0(x,x') + G_1(x,x')$ where $G_0$ is a smooth two point function which (when we restore $\hbar$) is of order $\hbar^0$ while $G_1$ differs from the Hadamard subtraction piece, $H_\ell$, of \eqref{Hadamard} by a smooth two-point function.   And when we further specialize to physical solutions \emph{to order $\hbar$} in Section \ref{Sect:1andhbar} then we will only demand that, $G_1$ differs \emph{to order $\hbar$} from the Hadamard subtraction piece, $H_\ell$, of \eqref{Hadamard} by a smooth two-point function.

We should bear in mind though that, when we say that a state, $\omega$, on a fixed curved spacetime has a certain dependence on $\hbar$, we have been taking for granted in the present section, that the only source of dependence on $\hbar$ is in the quantum field theory and that the background geometry does not involve $\hbar$.  However, when we say that a state, $\omega$, that figures in a (physical) solution (see Section \ref{Sect:SSC}) $(({\cal M}, g_{ab}), \omega)$, to semiclassical gravity  has a two-point function which is Hadamard, then  -- even if, for example, we were to choose the initial geometry, $(g_{ij}, \dot g_{ij})$, on an initial surface to be independent of $\hbar$ --  the relevant Hadamard subtraction piece, $H_\ell(x,x')$, will no longer simply be proportional to $\hbar$ but will have inherited a more complicated dependence on $\hbar$ due to the fact that the spacetime metric, $g_{ab}$, of the Cauchy development of our initial data will have acquired a nontrivial dependence on $\hbar$.  (And anyway we shall want to develop a theory of physical solutions where, as we shall discuss in Section \ref{Sect:SSC}, one is free to specify an arbitrary $\hbar$ dependence for initial data pairs $(g_{ij}, \dot g_{ij})$ provided only that data is then smooth as a function of $\hbar$ and of coordinates in any coordinate chart of our initial surface in neighbourhoods of arguments, $(x_1, x_2, x_3, \hbar)$, for which $\hbar = 0$. 

For states in the above general class (and whether we are in a fixed background or semiclassical context), we may refine the definition (\eqref{Tabren}, \eqref{semi-hbar}) of the right hand side of \eqref{semi-simple} by writing it as the sum 
\begin{equation}
\label{semi-refine}
\omega(T_{ab}^{\rm ren} ) := \omega(T_{ab})^{[0]} + \omega(T_{ab})^{[1]}
\end{equation} 
of a piece, $\omega(T_{ab})^{[0]}$, of order $\hbar^0$ defined to be $\lim_{x' \to x} \mathscr{T}_{ab}(x, x') G_0(x, x')$ and a piece, $\omega(T_{ab})^{[1]}$, of order $\hbar$, defined as in \eqref{semi-hbar} with $\omega(\phi(x)\phi(x'))$ replaced by $G_1(x,x')$.

\section{\label{Sect:SSC} Semiclassical gravity in the absence of quantum state collapses: physical solutions, the surface Hadamard condition and some initial value conjectures}

We may now formalize our opening informal definition in the introduction by defining a \emph{solution to semiclassical gravity} as consisting of a pair $(({\cal M}, g_{ab}), \omega)$ where $({\cal M}, g_{ab})$ is a ($C^\infty$) globally hyperbolic spacetime and $\omega$ a Hadamard state on the *-algebra $\mathscr{A}({\cal M}, g_{ab})$ such that the equation \eqref{semi-simple} holds with $\omega(T_{ab}^{\rm ren})$ as in Equation \eqref{semi-hbar}.   We remark that this is essentially the same thing as what was referred to in \cite{DiezTejedor:2011ci} as a `semiclassical self-consistent configuration' or `SSC'.  And we refer to that paper for further general discussion of solutions to semiclassical gravity.    

To define physical solutions, as discussed in the introduction, we must think of $g_{ab}$ and $\omega$ and $\omega(T^{\rm ren}_{ab})$ as functions of $\hbar$  -- as well, in the case of $g_{ab}$ and $\omega(T^{\rm ren}_{ab})$, as of coordinates in any chart on ${\cal M}$.   We may then formalize the notion of a \emph{physical solution}, as a solution, $(({\cal M}, g_{ab}), \omega)$, of \eqref{semi-simple} for which $g_{ab}$ and $\omega$ are smooth as a function of $\hbar$ and of coordinates in any coordinate chart of $\cal M$ in neighbourhoods of arguments, $(x_0, x_1, x_2, x_3, \hbar)$, for which $\hbar = 0$.    For our general class of states discussed in Section \ref{Sect:Dephbar})  we recall that $\omega(T^{{\rm ren}}_{ab})$ will arise as the sum of a piece, $\omega(T_{ab})^{[0]}$, of order $\hbar^0$ and a piece, $\omega(T_{ab})^{[1]}$, of order $\hbar$, we will thus have, in particular, that a physical solution will be approximated by -- i.e.\ will hold up to corrections of order $\hbar$ and higher to -- a solution to the equation 
\begin{equation}
\label{physsolapprox}
G_{ab} + \Lambda g_{ab} = 8\pi G_{\rm N}\omega(T_{ab}^{[0]}).
\end{equation}
which resembles a classical Einstein equation and in particular has no `higher derivative' terms.  (We shall discuss this order $\hbar^0$ approximation further in Section \ref{Sect:1andhbar}.)

Next we aim to discuss the initial value problem, both for unrestricted solutions to \eqref{semi-simple} and also for physical solutions. 

In view of (a) the well-posedness of the initial value problem for classical General Relativity (see e.g.\ \cite[Chapter 10]{Wald})  and (b) the fact that, in the quantum field theory of \eqref{KG} in a fixed curved spacetime $({\cal M}, g_{ab})$, an initial state, $\omega_{\cal C}$ on the CCR algebra, $\mathscr{A}({\cal C}, ^3\!g)$, of a Cauchy surface, $\cal C$, will determine a state $\omega = \omega_{\cal C}\circ {\rm iso}_{\cal C}^{-1}$ such that, informally, $\omega_{\cal C}$ is the `restriction' of $\omega$ to the initial surface, $\cal C$, it seems reasonable to expect that the system consisting of \eqref{semi-simple} and \eqref{KG} will have an initial value formulation in the sense that prescription of suitable classical gravitational data on $\cal C$ together with a suitable initial state, $\omega_{\cal C}$ on the CCR algebra, $\mathscr{A}({\cal C}, ^3\!g)$, of a 3-surface, $\cal C$ will determine a unique solution $(({\cal M}, g_{ab}), \omega)$ to the semiclassical gravity equations \eqref{semi-simple}, \eqref{KG} such that $\cal C$ is a Cauchy surface in $\cal M$ and such that the classical gravitational data will arise as the restriction of $g_{ab}$ and a suitable number of its derivatives to $\cal C$ and the state $\omega$ will restrict, in the sense of (b) above, to the initial state $\omega_{\cal C}$.

As for the suitable classical gravitational data, as we already mentioned in Section \ref{Sect:star},  the semiclassical Einstein equations have the character of fourth order equations in the sense that, for general solutions, we expect to need to specify on our initial surface, not only the metric $g_{ij}$ and the extrinsic curvature, $K_{ij}$ -- which is the same thing as (half) the restriction to our surface of the metric's time-derivative, $\dot g_{ij}$ --  but also to need to specify the restrictions to the surface of the second and third time derivatives, $g^{(2)}_{ij}$ and $g^{(3)}_{ij}$.

Amongst the conditions that will render a set of initial data (consisting of a quadruple $(g_{ij}, \dot g_{ij}, g^{(2)}_{ij}, g^{(3)}_{ij})$, of classical gravitational initial data together with a quantum state, $\omega_{\cal C}$, on the initial CCR algebra $\mathscr{A}({\cal C}, ^3\!g)$) suitable, we anticipate that the constraint equations will need to be satisfied.
I.e.\  the $0a$ components of \eqref{semi-simple} with right hand side given by the $\omega(T_{0a}^{\rm ren}(x) )$ of \eqref{semi-hbar} 
(See e.g.\ \cite[Section 2.2]{Juarez-Aubry:2021abq}).   Or rather, the result 
\begin{subequations}  
\label{GravConstraints}
\begin{align}
^{(3)}\!R + K^2-K^{ij}K_{ij} + 2\Lambda  & = 16G_{\rm N}\pi (\omega_{\cal C}(T^{{(\ell)}\rm ren\, 00}) +
\hbar(\hat\alpha_3 I^{00} + \hat\alpha_4 J^{00})), \\
D^j(K_{ij}-K g_{ij}) & =-8\pi G_{\rm N} (\omega_{\cal C}({T^{(\ell){\rm ren,0}}_{\ \ \ \ \ \ \ \ \, i}}) +
\hbar(\hat\alpha_3 I^0_{\ i} + \hat\alpha_4 J^0_{\ i})) 
\end{align}
\end{subequations}
of re-expressing those equations in terms of initial data.  Here, $K_{ij}$ is identified with half the piece of initial data, $\dot g_{ij}$, $D^j$ denotes the covariant derivative with respect to the metric connection for $g_{ij}$, while the definition (see Equation \eqref{Tabren}) of $\omega(T_{0a}^{\rm ren}(x))$ is assumed to have been rewritten, as in Equation \eqref{surfHad}, in terms of $\omega_{\cal C}$ and the metric, $g_{ij}$, and its first three time derivatives, $\dot g_{ij}, g^{(2)}_{ij}, g^{(3)}_{ij}$ in Gaussian normal coordinates).

As far as physical solutions are concerned, we make the conjecture (to be clarified further in Conjecture \ref{Conj:PhysSurf}):

\medskip

\begin{conj}
\label{Conj:Phys}
Instead of our above expectation for the initial value problem for general solutions, for a physical solution, specifying $g_{ij}$ and $K_{ij}$ ($=\frac{1}{2}\dot g_{ij}$) will suffice and moreover $g^{(2)}_{ij}$ and $g^{(3)}_{ij}$ will be determined by $g_{ij}$ and $K_{ij}$.  Here, we have of course, though, to think of the initial data as functions of $\hbar$ as well as of coordinates on the initial surface, $\cal C$, which are smooth as a function of $\hbar$ and of coordinates in any coordinate chart of our initial surface in neighborhoods of arguments, $(x_1, x_2, x_3, \hbar)$, for which $\hbar = 0$.  Also, and in consequence, in place of the above expectation for the constraint equations for general solutions, we expect that, for physical solutions, the constraints may be considered to be equations involving only $g_{ij}$ and $K_{ij}$ as in classical general relativity (together with $\omega_{\cal C}$).
\end{conj}

\medskip

To see why this conjecture is reasonable, consider, for example (see also the related discussion in \cite{Parker:1993dk}, the simple linear fourth order ordinary differential equation 
\begin{equation}
\label{4ordode}
\left(\hbar\frac{d^4}{dt^4} + \frac{d^2}{dt^2} + m^2\right)u=0,
\end{equation}
where $\hbar$ stands for a parameter and, analogously to our definition of physical solutions for \eqref{semi-simple}, let us define physical solutions for this equation to also be those that are  differentiable in $\hbar$ at $\hbar=0$.   Equation \eqref{4ordode} has the general (complex-valued) solution
\[
u = A e^{i\omega_{\mathrm{\mathrm{phys}}}t} + B e^{-i\omega_{\mathrm{\mathrm{phys}}}t} + C e^{i\omega_{\mathrm{\mathrm{unphys}}}t} + D e^{-i\omega_{\mathrm{\mathrm{unphys}}}t}
\]
where
\[
\omega_{\mathrm{phys}} = \sqrt{\frac{1 - \sqrt{1 - 4\hbar m^2}}{2\hbar}}= m + O(\hbar), \quad 
\omega_{\mathrm{unphys}} = \sqrt{\frac{1 + \sqrt{1 - 4\hbar m^2}}{2\hbar}} = 1/\sqrt{\hbar} + O(1)
\]
It is easy to see from this that, while one needs to set Cauchy data $h=u(0), k= \dot u(0), l=\ddot u(0), m=\dddot u(0)$ at, say, $u=0$ to determine a general solution, for a physical solution we must have
$l = - \omega_{\mathrm{phys}}^2 h, m = - \omega_{\mathrm{phys}}^2 k$ in analogy with our conjecture that   
$g^{(2)}_{ij}$ and $g^{(3)}_{ij}$ are determined by $g_{ij}$ and $\dot g_{ij}$.   Let us also notice that $l$ and $k$ differ by terms of $O(\hbar)$ from the values they would have 
(i.e.\  $l= - m^2 h$, \, $m = -m^2 k$) were we to solve the `classical' equation $(\frac{d^2}{dt^2} + m^2)u=0$ with initial data $h=u(0), k= \dot u(0)$, analogously to \eqref{physsolapprox}.

The question of what constitutes a suitable initial quantum state (for either general or physical solutions) is a more complicated issue and we turn to discuss it next.

For general solutions, we seek a condition -- we shall call it the \emph{surface Hadamard condition} -- which depends on our choice of initial gravitational data  $g_{ij}$, $\dot g_{ij}$, $g^{(2)}_{ij}$ and $g^{(3)}_{ij}$ on a surface $\cal C$, which will ensure that an initial state, $\omega_{\cal C}$, on $\cal C$ which satisfies it could arise, together with that classical gravitational data, as the restrictions to $\cal C$ of the (Hadamard) state $\omega$ and the metric $g_{ab}$ and its first 3 time derivatives, of a semiclassical solution, $(({\cal M}, g_{ab}), \omega)$, on a spacetime which has $\cal C$ as a Cauchy surface.  (Here when we say that $\omega_{\cal C}$ is the restriction of $\omega$,  we mean that $\omega\circ {\rm iso}_{\cal C} = \omega_{\cal C}$.)   Let us note here that we anyway at least need our initial state to be sufficently approximately Hadamard in the sense that, when we replace $H_\ell$ by the $H_\ell^{\rm DF}$ of Section \ref{Sect:star}, the limit in \eqref{Tabren} exists.   However (see Footnote \ref{ftnt:smooth}) to have $C^\infty$ solutions we expect that we will actually need our surface Hadamard condition to reflect the full strength of the full Hadamard condition. For physical solutions see Conjecture \ref{Conj:PhysSurf} below.

\subsection{\label{Sect:YukSurf} The scalar Yukawa case}

We shall first address the somewhat more straightforward analogous question of finding a suitable `surface Hadamard condition' for initial states on a constant time surface, $\cal C$ (which in this model we take to be a flat spacelike Cauchy surface in Minkowski space) in our semiclassical scalar Yukawa model, described by the equations \eqref{psieq}, \eqref{phieq}.

There are obvious counterparts to the field *-algebra, $\mathscr{A}({\cal M}, g_{ab})$, and the CCR algebra, $\mathscr{A}({\cal C}, ^3\!g)$, of Section \ref{Sect:star} and also for the isomorphism ${\rm iso}_{\cal C}$ of that section for the equation \eqref{phieq} and we shall use the same symbol for the latter.   The Hadamard condition on the two-point function for this model takes the form
\begin{align}
\omega(\phi(x) \phi(x')) = \frac{1}{8\pi^2} \left( \frac{1}{\Sigma(x, x')} +  V(x, x') \ln \left(\frac{\Sigma(x, x')}{\ell^2} \right) + W_{{\ell} }^{(\omega)}(x, x') \right)
\label{HadamardYukawa}
\end{align}
and we note also that since our spacetime and our initial surface, $\cal C$, are both flat, the squared geodesic distance $\sigma(x,x')$ between two points, $x$, $x'$ in $\cal C$ will equal $\Sigma(x,x')$ and also successive time derivatives of $\Sigma(x,x')$ restricted to $\cal C$ will, likewise be readily calculable.  (For an example of how such derivatives enter, see Equations \eqref{dtHell}, \eqref{dtprimeHell} below.)   By the techniques in \cite{DeWittBrehme,Decanini:2005eg} (or by equating the terms in \cite[Equations (108), (109)]{Decanini:2005eg} that arise from the $\xi R$ term there with those arising from the $2\lambda$ coefficient of $\chi$ in \eqref{phieq} while setting the other curvature terms to zero) we find that $V$ has a covariant Taylor expansion that begins as (see Footnote \ref{ftnt:covTay})
\begin{equation} 
\label{VYukcovTay}
V = \frac{m^2}{2} + \lambda \chi - \frac{\lambda}{2}\nabla_a \chi \Sigma^{;a} + \left(\frac{\lambda^2}{6}\nabla_a\nabla_b \chi + \left(\frac{(m^2+2\lambda \chi)^2}{16} -\frac{\lambda\Box \chi}{24}\right)\eta_{ab}\right)  \Sigma^{;a}\Sigma^{;b} + \ldots \, .
\end{equation}
To make sense of the `$\omega(\phi(x)^2)$' on the right hand side of \eqref{psieq}, we need, of course, to replace it by a renormalized $\omega(\phi(x)^{2 \, \mathrm{ren}})$, which (rather more simply than \eqref{Tabren}) we shall define by
\begin{equation}
\label{phisqren}
\omega(\phi(x)^{2 \, {\rm ren}}) = \lim_{x'\rightarrow x}(\omega(\phi(x)\phi(x')) - 
H_\ell(x,x'))
\end{equation}
where 
\begin{align}
H_\ell = \frac{1}{8\pi^2} \left( \frac{1}{\Sigma(x, x')} +  V(x, x') \ln \left(\frac{\Sigma(x, x')}{\ell^2} \right) \right)
\label{HellYukawa}
\end{align}

We note: (A) For the purpose of defining $\omega(\phi(x)^{2 \, {\rm ren}})$, we may approximate $H_\ell$ (analogously to the way we pointed out that the $H_\ell$ of Section \ref{Sect:star} may be approximated by $H^{\rm DF}_\ell$  for the purpose of defining $\omega(T_{ab}^{\rm ren})$) by truncating the covariant Taylor expansion \eqref{VYukcovTay} at its zeroth order term ($m^2/2 + \lambda \chi$) since the higher order terms in the expansion of $H_\ell$ obviously tend to zero in the $x' \rightarrow x$ limit. (B) It is easy to check that an alternative formula to \eqref{phisqren} is to be had by restricting the $x'$ in the limit to lie in the same constant time surface -- let us call it again ${\cal C}$ -- as $x$ and replacing $\omega$ by its restriction, $\omega_{\cal C}$, to that surface.   That is, we have, for $x$, $x'$ confined to our constant time surface, $\cal C$,
\begin{equation}
\label{phisqren3d}
\omega_{\cal C}(\phi(x)^{2 \, {\rm ren}}) = \lim_{x'\rightarrow x}(\omega_{\cal C}(\varphi(x)\varphi(x')) - H_\ell(x,x')),
\end{equation} 
(Here, and below, we let $\varphi$ and $\pi$, respectively, denote the restriction of $\phi$ and $\dot\phi$ to our initial surface.)    (C) In consequence of (A), changing the length scale, $\ell$, in the logarithmic term in $H_\ell$ can be compensated by a renormalization of the constants $\lambda$ and $\mu$ in Equations \eqref{psieq}.  (D) The above definition amounts to equating $\omega(\phi(x)^{2 \, {\rm ren}})$ with the coincidence limit of the $W_{{\ell} }^{(\omega)}(x, x'))$ term of \ref{HadamardYukawa} (for the chosen value of the length $\ell$).   However, just as for the gravitational case (see again Footnote \ref{ftnt:smooth}) all the terms of Equation \eqref{VYukcovTay} are needed to specify what we mean by a Hadamard state.  

So, to have a suitable notion of `surface Hadamard', we presumably need to know the values of all the time derivatives of $\chi$ on our initial constant time surface, $\cal C$.  If only we were given all of those, then we could compute $V$ using Equation \eqref{VYukcovTay} and hence, using Equation \eqref{HellYukawa}, compute $H_\ell$ as well as $\partial_t H_\ell$, $\partial_{t'} H_\ell$ and $\partial_t\partial_{t'}H_\ell$ for pairs of points, $x$ and $x'$ in $\cal C$.   We would then be able to give the following preliminary definition of a surface Hadamard state.  (We call it `preliminary' because it requires that all the time derivatives, $\chi$, $\dot \chi $, $\chi^{(2)}$, $\dots$, of $\chi$ be given whereas we are ultimately aiming for a definition that only requires the classical initial data, $\chi, \dot \chi$, to be given.) 

\begin{defn}
\label{Def:YukawaSurfPrelim}
An initial state $\omega_{\cal C}$ on a constant time surface $\cal C$ equipped with the restrictions to $\cal C$ of the function $\chi$ and all its time derivatives, $\dot \chi$, $\chi^{(2)}$, $\dots$ is said to be a \emph{preliminary surface Hadamard state} for those values $\chi, \dot \chi , \chi^{(2)}, \dots$ if, for all $x$ and $x'$ in $\cal C$, the two point functions 
$\omega_{\cal C}(\varphi(x)\varphi(x')), \omega_{\cal C}(\pi(x)\varphi(x')) (=\omega_{\cal C}(\varphi(x)\pi(x'))), \omega_{\cal C}(\pi(x)\pi(x'))$  differ from $H_\ell, \partial_t H_\ell, \partial_{t'} H_\ell$ and $\partial_t\partial_{t'}H_\ell$ respectively by smooth two point functions on $\cal C$, where the latter quantities are defined in terms of $\chi, \dot \chi, \chi^{(2)}, \dots$ in the way explained above.
\end{defn}

The problem we face is that only $\chi$ and its first time derivative, $\dot\chi$, are given to us as part of our initial data.  However, we may proceed iteratively, using the equation \eqref{psieq} and its successive time derivatives, on our initial surface, to find the values of the successive time derivatives of $\chi$ restricted to our initial surface.   As it turns out, as a byproduct of this iterative process, we generate a sequence of conditions on the initial state, $\omega_{\cal C}$, of a set of initial data, $\lbrace\chi, \dot \chi , \omega_{\cal C}\rbrace$ -- specifically, conditions on the three surface two-point functions,
$\omega_{\cal C}(\varphi(x)\varphi(x'))$, $\omega_{\cal C}(\varphi(x)\pi(x'))$  and $\omega_{\cal C}(\pi(x)\pi(x'))$ which must be satisfied for the successive time derivatives to have finite values. (See after Equation \eqref{four2points}.)  To illustrate how this works, at the first stage of this process, we have, directly from \eqref{psieq}:
\begin{equation}
\label{ddotaEq}
\ddot \chi = \nabla^2 \chi - M^2 \chi + \mu - \lambda\omega_{\cal C}(\varphi(x)^{2 \, {\rm ren}}),
\end{equation}
and this is well-defined provided the limit \eqref{phisqren3d}, which defines $\omega_{\cal C}(\varphi(x)^{2 \, {\rm ren}})$, exists.

At the second stage, we find
\begin{equation}
\label{3dotaEq}
\chi^{(3)} = \nabla^2 \dot \chi - M^2 \dot \chi - \lambda \lim_{x'\rightarrow x}(\omega_{\cal C}(\pi(x)\varphi(x')) - \partial_t H_\ell(x,x') + \omega_{\cal C}(\varphi(x)\pi(x')) - \partial_{t'} H_\ell(x,x')).
\end{equation}
By \eqref{HellYukawa} and \eqref{VYukcovTay}, we have
\begin{equation}
\label{dtHell}
8\pi^2\partial_t H_\ell = -\frac{1}{\Sigma^2}\Sigma_{;0} + \left( \lambda \dot \chi - \frac{\lambda}{2}\nabla_b \chi \Sigma^{;b}_{\ ;0} + \dots \right)\ln\Sigma + \left(\dots\right)\frac{1}{\Sigma}\Sigma_{;0} + \dots
\end{equation}
(where the dots indicate terms that will disappear in the limit)
\[
= \frac{\lambda\dot \chi}{2}\ln\Sigma + \hbox{terms that will disappear in the limit},
\]
and
\begin{equation}
\label{dtprimeHell}
8\pi^2\partial_t' H_\ell = -\frac{1}{\Sigma^2}\Sigma_{;0} + \left(-\frac{\lambda}{2}\nabla_b \chi \Sigma^{;b}_{\ ;0'} + \dots \right)\ln\Sigma +\left(\dots\right)\frac{1}{\Sigma}\Sigma_{;0} + \dots
\end{equation}
\[
= -\frac{\lambda\dot \chi}{2}\ln\Sigma + \hbox{terms that will disappear in the limit}.
\]

Here, we have used the fact that, in the limit $x' \rightarrow x$, $\Sigma^{;b}_{\ ;a}$ will tend to $\delta^b_a$ and $\Sigma^{;b}_{\ ;a'}$ (the bitensor resulting from acting on $\Sigma$ with a covariant derivative (with raised index $b$) in its $x$ argument and with a covariant derivative of its $x'$ argument) to $-\delta^b_a$ -- in the case $a = a' = 0$.
Thus $\chi^{(3)}$ is well defined in terms of $\dot \chi$ (which is part of the initial data) provided the state satisfies the conditions
\[
\lim_{x'\rightarrow x}\left(\omega_{\cal C}(\pi(x)\varphi(x')) - \frac{\lambda\dot \chi }{16\pi^2}\ln\Sigma\right) \ \hbox{and} \ \lim_{x'\rightarrow x}\left(\omega_{\cal C}(\varphi(x)\pi(x')) + \frac{\lambda\dot \chi}{16\pi^2}\ln\Sigma\right) \quad \hbox{exist}.
\]
At the third stage, we have
\begin{equation}
\label{3rdstage}
\chi^{(4)} = \nabla^2 \ddot \chi - M^2 \ddot \chi 
\end{equation}
\[
- \lambda \lim_{x'\rightarrow x}(\omega_{\cal C}(\ddot\varphi(x)\varphi(x')) - \partial_t^2 H_\ell(x,x') + \omega_{\cal C}(\pi(x)\pi(x')) - \partial_t\partial_{t'} H_\ell(x,x') + \omega_{\cal C}(\varphi(x)\ddot\varphi(x')) - \partial_{t'}^2 H_\ell(x,x')).
\]

One may then use \eqref{phieq} to replace $\ddot\phi$ above by $\nabla^2\phi -(m^2 + 2\lambda \chi)\phi$ and again replace $\partial_t^2 H_\ell(x,x')$, $\partial_t\partial_{t'} H_\ell(x,x')$ and $\partial_{t'}^2 H_\ell(x,x')$ by quantities which differ from them by terms that tend to zero in the limit and which only depend on $\chi$, $\dot\chi$ and $\ddot\chi$ and, in this way, one again obtains further conditions on the three surface two-point functions, $\omega_{\cal C}(\varphi(x)\varphi(x'))$, $\omega_{\cal C}(\pi(x)\varphi(x'))$, and $\omega_{\cal C}(\pi(x)\pi(x'))$ of $\omega_{\cal C}$ which also depend only on $\chi$, $\dot \chi $ and $\ddot \chi $ for the limit in \eqref{3rdstage} to exist so that $\chi^{(4)}$ has a well-defined value.

It seems clear that one may continue this process:

\begin{conj}
\label{Conj:YukIt}
At the $k$th stage, we will find conditions on (the above three surface two point functions of) the initial state, $\omega_{\cal C}$, which depend only on the values of $\chi$ and of its derivatives up to the $(k-2)$nd derivative, which were found at earlier stages and which thereby determine the $(k+1)$st time derivative of $\chi$.
\end{conj}

Assuming this conjecture to hold, we then define a surface Hadamard state as follows:

\begin{defn}
\label{Def:YukawaSurf}
We say that an initial state, $\omega_{\cal C}$, on an initial surface $\cal C$ satisfies
the \emph{surface Hadamard condition} for a classical initial data pair $(\chi, \dot \chi)$ if (a) it satisfies all the conditions on $\omega_{\cal C}$ that arise at all the stages of this iterative process starting with the given values, $\chi$ and $\dot \chi$ and (b) it is preliminary surface Hadamard in the sense of Definition \ref{Def:YukawaSurfPrelim} for the full collection of values, $\chi, \dot \chi, \chi^{(2)}, \dots $, generated by the above iterative process.
\end{defn}

Of course, once one knows the full collection of values $\chi, \dot \chi, \chi^{(2)}, \dots $, (b) here will entail (a).  However since (a) is required for the values $\chi, \dot \chi, \chi^{(2)}, \dots $ to be known in terms of the initial data $\chi, \dot \chi$, it seems we cannot shorten this definition. 

\begin{rem}
\label{Rem:SurfHad}
We point out that the notion of `preliminary surface Hadamard' could be related to the notion of `Hadamard' in a 4-dimensional context for any $\phi$ field satisfying Equation \eqref{phieq} for any background $\chi$ field in the sense that the restriction of a Hadamard state for that background to an initial surface would be preliminary surface Hadamard when $\chi, \dot\chi, \dots$ in Definition \ref{Def:YukawaSurfPrelim} are taken to be the restrictions to the surface of $\chi, \dot\chi, \dots$ for the given background field $\chi$.    On the other hand the notion of `surface Hadamard' involves both the equations \eqref{psieq} and \eqref{phieq}.   When we generalize these notions to semiclassical gravity in Section \ref{Sect:SufHadSemiclass} the corresponding state of affairs may be expressed, informally, by saying that `preliminary surface Hadamard' may be thought of as a quantum field theory in curved spacetime notion, while `surface Hadamard' is a notion that only makes sense for semiclassical gravity.
\end{rem}

Now we have a definition for a surface Hadamard initial state. We may state the further conjecture that the initial value problem is well posed for this model in the following sense:
\begin{conj}
\label{Conj:YukawaIV}
Given an initial constant time surface, $\cal C$, and any pair of classical initial data, $(\chi_{\rm in}, \dot \chi_{\rm in})$, on $\cal C$ and given an initial state, $\omega_{\cal C}$, on the CCR algebra of $\cal C$, which is surface Hadamard for that classical initial data, then there will be a unique solution, $(\chi, \omega)$ ($\chi$ a scalar function on Minkowski space, $\omega$ a state on the *-algebra of the quantum equation \eqref{phieq}) to \eqref{ScalScalar} such that $\chi $ and $\dot \chi$ coincide with $\chi_{\rm in}$ and $\dot \chi_{\rm in}$ when restricted to $\cal C$ and $\omega$ coincides with $\omega_{\cal C}$ in the sense that $\omega\circ {\rm iso}_{\cal C} = \omega_{\cal C}$.
\end{conj}

\subsection{\label{Sect:ElecSurf} Semiclassical electrodynamics}

As in the scalar Yukawa case, there are, again, obvious counterparts to the field *-algebra, $\mathscr{A}({\cal M}, g_{ab})$, and the CCR algebra, $\mathscr{A}({\cal C}, ^3\!g)$, of Section \ref{Sect:star} for the equation \eqref{chphieq},\footnote{\label{ftnt:nonHerm} Note that the field algebra and CCR algebra are now for the charged (i.e.\ non-Hermitian) scalar field $\phi$ of Equation \eqref{chphieq} and will depend on the choice of gauge but (formal) quantities, such as $\phi^*\phi$ and $j^a$, which are formally invariant under gauge transformations ($A_a \rightarrow A_a + \partial_a\Lambda$, $\phi \rightarrow e^{i\Lambda}\phi$) are independent of that choice.   Note also that, in place of the three quantities $\omega_{\cal C}(\varphi(x)\varphi(x'))$, $\omega_{\cal C}(\pi(x)\varphi(x'))$, $\omega_{\cal C}(\pi(x)\pi(x'))$ in the case of a neutral scalar field, the initial data for the two-point function in the case of a charged scalar field will consist of the six quantities $\omega_{\cal C}(\varphi(x)\varphi(x'))$, $\omega_{\cal C}(\pi(x)\varphi(x'))$, $\omega_{\cal C}(\pi(x)\pi(x'))$, $\omega_{\cal C}(\varphi(x)\varphi^*(x'))$, $\omega_{\cal C}(\pi(x)\varphi^*(x'))$, $\omega_{\cal C}(\pi(x)\pi^*(x'))$ (where $\pi$ now means $\dot\phi^*+ieA_0\phi^*$)} where now $g_{ab}$ is the Minkowski metric $\eta_{ab}$, and also for the isomorphism ${\rm iso}_{\cal C}$ and we shall again use the same symbol for the latter. We are interested in the semiclassical system \eqref{SemiElectro} in Minkowski spacetime presented in the introduction.

The suitable notion of Hadamard state for charged scalars has been thoroughly studied by Balakumar and Winstanley in \cite{BalaWinst}. The singular structure of the Wightman function in a Hadamard state is of the form
\begin{align}
H_\ell = \frac{1}{8\pi^2} \left( \frac{U(x,x')}{\Sigma(x, x')} +  V(x, x') \ln \left(\frac{\Sigma(x, x')}{\ell^2} \right) \right),
\label{HellElectro}
\end{align}
where now the coefficients $U$ and $V$ obey Hadamard recursion relations adapted to Equation\eqref{chphieq}. We refer to \cite{BalaWinst} for details. One has
\begin{subequations}
\begin{align}
U(x,x') & = 1 + i e A_a \Sigma^{;a} +\left(- \frac{1}{2} i e A_{(a,b)} - \frac{1}{2} e^2 A_a A_b \right) \Sigma^{;a} \Sigma^{;b} \nonumber \\
& + \frac{i e}{6} D_{(a}D_b A_{c)} \Sigma^{;a} \Sigma^{;b} \Sigma^{;c} + O(\Sigma^2), \\
V(x,x') & = \frac{m^2}{2} + \frac{i q m^2 }{2} A_a \Sigma^{;a} - \frac{i q}{12} \partial^b F_{ba} \Sigma^{;a} + \left( - \frac{i q m^2}{4} D_{(a}A_{b)} \right. \nonumber \\
& \left. - \frac{q^2}{24} F^{c}{}_a F_{bc} -\frac{q^2}{12} A_{(a} \partial^c F_{b)c} - \frac{i q}{24} \partial_{(a} \partial^c F_{b)c} + \frac{m^4}{16}  \eta_{ab} \right) \Sigma^{;a}\Sigma^{;b} + O(\Sigma^{3/2}).
\end{align}
\end{subequations}
where $D_{a} := \partial_a - i e A_a$. The expectation value of the renormalized current in a state $\omega$ is defined (cf.\ Equation \eqref{ja}) by
\begin{align}
\omega(j_a^{\rm ren})
& = -ie \lim_{x'\to x}(\eta_{a}{}^{a'}(\partial_{a'} - i e A_{a'})  - \partial_a - i e A_a) (\omega(\phi^*(x) \phi(x')) - H_\ell(x,x')),
\end{align}
where $A_{a'}$ is evaluated at $x'$ and primed indices act on the spacetime point $x'$, whereas $A_a$ is evaluated at $x$  and unprimed indices act on the spacetime point $x$. 

We expect that the initial data for the system \eqref{SemiElectro} will consist of the electric and magnetic field on the $t = 0$ initial surface, together with a state defined on the CCR algebra of the initial surface. The data for the electric and magnetic field can be equivalently specified, once a gauge is chosen, by $A^a$ and its first time derivative at $t = 0$.

To keep the discussion general, we do not choose a gauge at the moment, although it is well known that to make sense of the initial value formulation of classical electrodynamics it is useful to work in the Coulomb gauge, and we will indeed make this gauge choice below for stating our conjectures.

Analogously to the
Yukawa model case, we have as a first step the condition
\begin{align}
-\ddot A_b + \nabla^2 A_b + \partial_b \dot A_0 - \partial_b \partial^i A_i = - 4 \pi \omega_\mathcal{C}(j_b^{\rm ren}),
\end{align}
where $\nabla^2$ is the flat-space Laplacian. This imposes the following conditions on initial data,
\begin{align}
\nabla^2 A_0 - \partial^i \dot A_i & = 4 \pi i e \lim_{x' \to x} \left( (\omega_\mathcal{C}(\varphi(x) \pi (x') ) - \partial_{t'} H_\ell(x,x')) - (\omega_\mathcal{C}(\pi(x) \varphi (x') ) - \partial_t H_\ell(x,x')) \right. \nonumber \\
& \left. -2 i e A_0 (\omega_\mathcal{C}(\varphi(x) \varphi (x') ) - H_\ell(x,x'))  \right),
\label{ElectroConstr1}
\end{align}
which is the Gauss constraint equation, together with
\begin{align}
-\ddot A_i + \nabla^2 A_i + \partial_i \dot A_0 - \partial_i \partial^j A_j = 4 \pi i e \lim_{x' \to x} \left( \eta_j{}^{j'}\partial_{j'} - \partial_j - 2 i e A_j  \right)\left(\omega_\mathcal{C}(\varphi(x) \varphi (x') ) - H_\ell(x,x')\right),
\label{ElectroDyn1}
\end{align}
which are the dynamical equations.

We proceed now as in the case of the Yukawa model above, considering only truncations of the subtraction piece. For Eq.\ \eqref{ElectroDyn1} we can see that the only terms contributing to the subtraction are
\begin{align}
H_\ell(x,x') &= \frac{1}{8\pi^2} \left( \frac{1 + i e A_a \Sigma^{;a} +\left(- \frac{1}{2} i e A_{(a,b)} - \frac{1}{2} e^2 A_a A_b \right) \Sigma^{;a} \Sigma^{;b}}{\Sigma(x, x')} +  \frac{m^2}{2} \ln \left(\frac{\Sigma(x, x')}{\ell^2} \right) \right) + \ldots
\end{align}
where the dots denote terms that vanish in the limit and that should be ignored at this stage. Thus, we can replace Eq. \eqref{ElectroDyn1} by
\begin{align}
& -\ddot A_i + \nabla^2 A_i + \partial_i \dot A_0 - \partial_i \partial^j A_j  = 4 \pi i e \lim_{x' \to x} \left( \eta_j{}^{j'}\partial_{j'} - \partial_j - 2 i e A_j  \right)\left[\omega_\mathcal{C}(\varphi(x) \varphi (x') ) \right. \nonumber \\
& \left. - \frac{1}{8\pi^2} \left( \frac{1 + i e A_a \Sigma^{;a} +\left(- \frac{1}{2} i e A_{(a,b)} - \frac{1}{2} e^2 A_a A_b \right) \Sigma^{;a} \Sigma^{;b}}{\Sigma(x, x')} +  \frac{m^2}{2} \ln \left(\frac{\Sigma(x, x')}{\ell^2} \right) \right) \right].
\label{ElectroDyn2}
\end{align}

For the constraint \eqref{ElectroConstr1} we have
\begin{subequations}
\begin{align}
H_\ell(x,x') &= \frac{1}{8\pi^2} \left( \frac{1 + i e A_a \Sigma^{;a} +\left(- \frac{1}{2} i e A_{(a,b)} - \frac{1}{2} e^2 A_a A_b \right) \Sigma^{;a} \Sigma^{;b}}{\Sigma(x, x')} +  \frac{m^2}{2} \ln \left(\frac{\Sigma(x, x')}{\ell^2} \right) \right) + \ldots \\
\partial_{t'} H(x,x') & =  \frac{1}{8\pi^2} \left( \frac{ \partial_{t'} U(x,x')}{\Sigma(x, x')} - \frac{ U(x,x')}{\Sigma(x, x')^2}  \partial_{t'} \Sigma(x,x') \right. \nonumber \\
& \left. +  \partial_{t'} V(x, x') \ln \left(\frac{\Sigma(x, x')}{\ell^2} \right) + \frac{V(x,x')}{\Sigma(x, x')} \partial_{t'} \Sigma(x,x') \right) + \ldots,\\ 
\partial_t H(x,x') & = \frac{1}{8\pi^2} \left( \frac{ \partial_t U(x,x')}{\Sigma(x, x')} - \frac{ U(x,x')}{\Sigma(x, x')^2}  \partial_t \Sigma(x,x') \right. \nonumber \\
& \left. +  \partial_t V(x, x') \ln \left(\frac{\Sigma(x, x')}{\ell^2} \right) + \frac{V(x,x')}{\Sigma(x, x')} \partial_t \Sigma(x,x') \right) + \ldots,
\label{ElectroConstraintHadSub}
\end{align}
\end{subequations}
where again the dots denote terms that vanish in the coincidence limit, and here
\begin{align}
\partial_t U & =  i e \partial_t A_a \Sigma^{;a} +\left(- \frac{1}{2} i e \partial_t A_{(a,b)} -  e^2 A_a \partial_t A_b \right) \Sigma^{;a} \Sigma^{;b} \nonumber \\
&  + i e A_a \partial_t \Sigma^{;a} +\left(- i e A_{(a,b)} - e^2 A_a A_b \right) \Sigma^{;a} \partial_t \Sigma^{;b} + \frac{i e}{2} D_{(a}D_b A_{c)} \Sigma^{;a} \Sigma^{;b} \partial_t \Sigma^{;c} 
+ \ldots, \\
\partial_{t'} U & =  i e A_a \partial_{t'} \Sigma^{;a} +\left(- i e A_{(a,b)} - e^2 A_a A_b \right) \Sigma^{;a} \partial_{t'} \Sigma^{;b} + \frac{i e}{2} D_{(a}D_b A_{c)} \Sigma^{;a} \Sigma^{;b} \partial_{t'} \Sigma^{;c} + \ldots, \\
V & = \frac{m^2}{2} + \frac{i q m^2 }{2} A_a \Sigma^{;a} - \frac{i q}{12} \partial^b F_{ba} \Sigma^{;a} + \ldots , \\
\partial_t V & = \frac{m^2}{2} + \frac{i q m^2 }{2}  A_a \partial_t \Sigma^{;a} - \frac{i q}{12} \partial^b  F_{ba} \partial_t \Sigma^{;a} + \ldots, \\
\partial_{t'} V & = \frac{m^2}{2} + \frac{i q m^2 }{2} A_a \partial_{t'}\Sigma^{;a} - \frac{i q}{12} \partial^b F_{ba} \partial_{t'}\Sigma^{;a} + \ldots. 
\label{Electro-UVtruncations}
\end{align}

Inserting the truncations \eqref{Electro-UVtruncations} into \eqref{ElectroConstraintHadSub} we obtain a condition analogous to \eqref{ElectroDyn2}. 

The process can now be iterated analogously as in the Yukawa case. However, a key difference in this example is that it involves `implicit equations'.  
By this we mean that already at the first step an appropriate truncation of $H_\ell(x,x')$ in the Hadamard subtraction requires control of second-order time derivatives of $A_a$. Thus, terms of the form $A_a^{(2)}$ appear both inside and outside the subtraction limits at the first stage of the tower of conditions that the data of semiclassical electrodynamics should satisfy. This situation is expected to continue at the successive stages of an iterative process, with the second step involving $A_a^{(3)}$, etc. We remark that this feature is shared by semiclassical gravity where, as we shall see, the first step also involves implicit equations  -- in the gravity case for the fourth time derivative of the spacetime metric on the initial surface. 

In view of the above, in the semiclassical electrodynamics case, in place of Conjecture \ref{Conj:YukIt} we make the following conjecture working in the Coulomb gauge:
\begin{conj}
\label{Conj:EMIt}
At the $k$th stage, we will find conditions on (the
six [see Footnote \ref{ftnt:nonHerm}] surface two point functions of) the initial state, $\omega_{\cal C}$, which depend on the values of $A_a$ and $\dot A_a$ (equivalently ${\bf B}$ and ${\bf E}$) and up to
the $(k+1)$st time-derivative of $A$ in the form of implicit equations. The implicit equations subsume the Gauss constraint equation, $\nabla_iE_i = 4\pi\omega_{\cal C}(j_0^{\rm ren})$) at the first stage and ensure the constraint conservation at the subsequent stages.
\end{conj}

Assuming this conjecture to hold, we can then define a notion of \emph{preliminary surface Hadamard state} analogous to Def. \ref{Def:YukawaSurfPrelim}, which will then allow us to define the notion of a surface Hadamard state in semiclassical electrodynamics as follows:

\begin{defn}
\label{Def:EMSurf}
We say that an initial state, $\omega_{\cal C}$, on an initial surface $\cal C$ satisfies
\emph{the surface Hadamard condition} for a classical initial data pair $(A_a, \dot A_a)$ (equivalently $({\bf B}, {\bf E})$) if (a) it satisfies all the conditions on $\omega_{\cal C}$ that arise at all the stages of this iterative process starting with the given values, $A_a$ and $\dot A_a$ and (b) it is preliminary surface Hadamard, now for the full collection of values, $A_a, \dot A_a, A_a^{(2)}, \dots $, generated by the above iterative process.
\end{defn}

See the clarifying Remark \ref{Rem:SurfHad} for the relation between the notions of `preliminary surface Hadamard' and `surface Hadamard'.

Next, we have a conjecture about the well-posedness of the initial value problem for \eqref{SemiElectro}, which it is helpful for this purpose to rewrite in the form of the semiclassical Maxwell equations 
\begin{subequations}
\label{Maxwell}
\begin{align}
{\bf\nabla}.{\bf E} =4\pi\omega(\rho),\label{Max1}\\
{\bf\nabla}.{\bf B}=0,\label{Max2}\\
{\bf\nabla}\times {\bf E} = -\frac{\partial{\bf B}}{\partial t},\label{Max3}\\
{\bf \nabla}\times {\bf B} = \frac{\partial{\bf E}}{\partial t}+ 4\pi\omega({\bf j})\label{Max4},
\end{align}
\end{subequations}
where ${\bf E} = -\nabla\Phi - \dot{\bf A}$ and ${\bf B} = {\rm curl}{\bf A}$ where $\Phi = A_0 \ (= - A^0)$.  Also $\rho= j_0 = - j^0$ and $\bf j$ is the 3-vector with components $j^i$

\begin{conj}
\label{Conj:EMIV}
Given initial electric and magnetic fields, ${\bf E}^{\rm in}$ and ${\bf B}^{\rm in}$, on an initial constant time surface, $\cal C$, and a surface Hadamard state, $\omega_{\cal C}$, on the CCR algebra for that surface, there will be unique  electric and magnetic fields, $\bf E$ and $\bf B$, with initial values ${\bf E}^{\rm in}$ and ${\bf B}^{\rm in}$ and a state, $\omega$, on the *-algebra for Equation \eqref{chphieq} satisfying the full set of equations \eqref{Maxwell} such that the restriction of $\omega$ to $\cal C$ is equal to $\omega_{\cal C}$ (strictly $\omega\circ {\rm iso}_{\cal C} = \omega_{\cal C}$)
\end{conj}

We leave open the question of whether surface Hadamard states exist for all pairs
${\bf E}^{\rm in}$ and ${\bf B}^{\rm in}$ -- cf.\ the remark after Conjecture \ref{Conj:GravityIV} in Section \ref{Sect:SufHadSemiclass}.  Part of the issue here is whether, for given $\bf E$ and $\bf B$, an initial quantum state, $\omega_{\cal C}$ can be found for the charged scalar field such that the Gauss constraint equation,  $\nabla_iE^{\rm in}_i = 4\pi\omega_{\cal C}(j_0^{\rm ren})$, will hold.

We remark that the system of equations \eqref{Maxwell} may be re-expressed in terms of the electric field and the magnetic vector potential, $A_i$ in Coulomb gauge, $\nabla_iA_i = 0$
\begin{subequations}
\label{ScalElecCoul}
\begin{align}
{\rm div}{\bf E} \ (= -{\bf\nabla}.{\bf\nabla}\Phi)=4\pi\omega(\rho), \label{Coulomb1}\\
\partial^a\partial_a{\bf A} = - 4\pi \omega({\bf j}) +\frac{\partial{\bf\nabla}\Phi}{\partial t}
\end{align}
\end{subequations}
where we recall $\Phi = A_0$ ($= -A^0$) and we can then fix the definitions of the field and CCR *-algebras by taking $A_a$ to be in this gauge.   In terms of this formulation, our conjecture becomes
\begin{conj}
Given initial electric field, ${\bf E}^{\rm in}$, and initial Coulomb-gauge magnetic vector potential, ${\bf A}^{\rm in}$, on an initial constant time surface, $\cal C$, and a surface Hadamard state, $\omega_{\cal C}$, on the CCR algebra for that surface, then there will be a unique electric field, $\bf E$, and a unique magnetic vector potential, $\bf A$, with initial values ${\bf E}^{\rm in}$ and ${\bf A}^{\rm in}$ and a state, $\omega$, on the *-algebra for Equation \eqref{chphieq} satisfying the full set of equations \eqref{ScalElecCoul} such that the restriction of $\omega$ to $\cal C$ is equal to $\omega_{\cal C}$ (strictly $\omega\circ {\rm iso}_{\cal C} = \omega_{\cal C}$) 
\end{conj}

\subsection{The surface Hadamard notion and initial value conjecture for semiclassical gravity}
\label{Sect:SufHadSemiclass}

Turning to the notion of `surface Hadamard' for general solutions to semiclassical gravity, let us first notice that, in order to rewrite the definition, Equation \eqref{Tabren}, of the renormalized stress energy tensor, as an equation involving only quantities on a given initial surface, $\cal C$, analogously to the way we rewrote Equation \eqref{phisqren} as Equation \eqref{phisqren3d} for the quantity $\omega_{\cal C}(\phi(x)^{2 \, {\rm ren}})$ for the scalar Yukawa model above, we may rewrite the differential operator, $\mathscr{T}_{ab}(x,x')$, which appears in Equation \ref{Tabren}, and is defined in Equation \eqref{TabPointSplit}, in the form
\begin{equation}
\label{4bits}
\mathscr{T}_{ab}(x,x') = \mathscr{T}^{(1)}_{ab}(x,x') + \mathscr{T}^{(2)}_{ab}(x,x')\partial_t
+ \mathscr{T}^{(3)}_{ab}(x,x')\partial_{t'} + \mathscr{T}^{(4)}_{ab}(x,x')\partial_t\partial_{t'}
\end{equation}
where we work in Gaussian normal coordinates for $\cal C$ and $\mathscr{T}^{(1)}_{ab}(x,x')$, $\mathscr{T}^{(2)}_{ab}(x,x')$, $\mathscr{T}^{(3)}_{ab}(x,x')$, $\mathscr{T}^{(4)}_{ab}(x,x')$ are bidifferential operators on ${\cal C}$.\footnote{We remark here that one might have expected that a term $\mathscr{T}^{(5)}_{ab}(x,x')\partial_t\partial_t$ might be needed.  However we note that the term of this form which arises from the term 
$- g_{a}\,^{a'} g_{b}\,^{b'} \nabla_{a'} \nabla_{b'}$ 
in Equation \eqref{TabPointSplit} will cancel with the term of this form which arises from the term $g_{ab} g^{c d}\nabla_c \nabla_d$ in \eqref{TabPointSplit} in the coincidence limit. This can be readily seen at the level of the classical stress-energy tensor, cf.\ eg.\ \eqref{TabClass}, and can be generalized to Equation \eqref{TabPointSplit} by noting that $\omega(\phi(x)\phi(x')) - H_\ell(x,x')$ is a smooth, symmetric bifunction, as we have discussed below Equation \eqref{TabPointSplit}.}

Thus we expect that for pairs of points, $x, x'$ in an initial surface, $\cal C$, 
we may replace the first term, $\mathscr{T}_{ab}(x,x') \left[\omega(\phi(x)\phi(x') )- H_\ell(x,x') \right]$ in \eqref{Tabren} by 
\begin{equation}
\label{surfHad}
\begin{split}
\mathscr{T}^{(1)}_{ab}(x,x')\left[\omega_{\cal C}(\varphi(x)\varphi(x') )- H_\ell(x,x') \right] + \mathscr{T}^{(2)}_{ab}(x,x')\left[\omega_{\cal C}(\pi(x)\varphi(x') ) - \partial_tH_\ell(x,x') \right] \\
+ \mathscr{T}^{(3)}_{ab}(x,x')\left[\omega_{\cal C}(\varphi(x)\pi(x') )- \partial_{t'}H_\ell(x,x') \right] + \mathscr{T}^{(4)}_{ab}(x,x')\left[\omega_{\cal C}(\pi(x)\pi(x') )- \partial_t\partial_{t'}H_\ell(x,x') \right]
\end{split}
\end{equation}
where $\mathscr{T}^{(1)}_{ab}(x,x')$, $\mathscr{T}^{(2)}_{ab}(x,x')$, 
$\mathscr{T}^{(3)}_{ab}(x,x')$,  $\mathscr{T}^{(4)}_{ab}(x,x')$ are as in \eqref{4bits}, and perform the resulting limit in $\cal C$.

For the above purpose of defining the right hand side of Einstein's equations, we may, as we remarked in Section \ref{Sect:star}, replace the exact Hadamard subtraction piece, $H_\ell$, by what we called there $H_\ell^{\rm DF}$.  But, just as in the scalar Yukawa or electodynamics models, we will also anyway need, for initial classical gravitational data,  $g_{ij}, \dot g_{ij}, g^{(2)}_{ij}, g^{(3)}_{ij}$, on an initial surface, $\cal C$, to be able to identify a set of `surface Hadamard' states which will have well-defined singular structures for the four unapproximated quantities $H_\ell, \partial_t H_\ell, \partial_{t'}H_\ell, \partial_t\partial_{t'}H_\ell$ on $\cal C$. This in turn will require it to be possible, for any such state, to infer the values of all the time derivatives (i.e.\ $g^{(n)}_{ij}$ for all $n$) of the metric on $\cal C$.   In either case (i.e.\ approximate subtraction piece $H_\ell^{\rm DF}$, or exact subtraction piece, $H_\ell$) this will require us to deal with two issues that did not arise in the scalar Yukawa model, although the second had a counterpart in semiclassical electrodynamics.

The first such issue is that, for a general initial surface, $\cal C$, the (squared) length, $\sigma(x,x')$, of a geodesic confined to $\cal C$ between a pair of nearby points $x, x'$ in $\cal C$ will no longer be the same as the (squared) geodesic distance, $\Sigma(x,x')$, between the same pair of points as measured in the spacetime, since a spacetime geodesic may leave the surface.  (The only exception would be when $\cal C$ is a surface of time symmetry -- i.e.\ a surface with everywhere zero extrinsic curvature.)\footnote{Were we only interested in a theory which predicts, say, the solution to the future of a Cauchy surface, ${\cal C}$ from knowledge of the solution to its past, then by restricting attention to initial Cauchy surfaces, ${\cal C}$ which are concave to the past, in the sense that any geodesic connecting any pair of points in the surface lies to the past of the surface, then we could obviously avoid this problem.  However for our purposes in Section \ref{Sect:collapse}, we need a theory of the initial value problem which predicts the future from data on a surface which has no past, in the sense that it is the result of a sudden jump in the geometry which happens on that surface.}    However, we may overcome this difficulty by expanding $\Sigma$ as well as $\partial_t\Sigma$, $\partial_{t'}\Sigma$ and $\partial_t\partial_{t'}\Sigma$ in \emph{3-dimensional} covariant Taylor series in $\sigma$.  Thus, for example we may write
\begin{equation}
\label{SigmaN}
\Sigma = \Sigma_0 + \Sigma_1^{i_1}\sigma_{;i_1} + \tfrac{1}{2} \Sigma_2^{i_1i_2}\sigma_{;i_1}\sigma_{;i_2} +\dots  \tfrac{1}{N!} \Sigma_N^{i_1\dots i_N}\sigma_{;i_1}\dots\sigma_{;i_N} + \dots
\end{equation}
where $\sigma_{;a}$ denotes the derivative of $\sigma$ with respect to $x^a$.   The first four terms of this are readily calculated and one finds
\begin{equation}
\label{Sigsig}
\Sigma = \sigma  + \tfrac{1}{24}K_{(ij}K_{kl)}\sigma_{;i}\sigma_{;j}\sigma_{;k}\sigma_{;l} + \dots
\end{equation}
where $K_{ij}$ is the extrinsic curvature of $\cal C$.    For the higher terms, and for the expansions of
$\partial_t\Sigma$, $\partial_{t'}\Sigma$ and $\partial_t\partial_{t'}\Sigma$, we refer to our upcoming paper \cite{companion}.

Now, if only we were given the initial values of all of the successive time-derivatives, $g^{(n)}$, of the metric on $\cal C$ then we would be able to calculate all the coefficients in the 4-dimensional covariant Taylor expansion \eqref{UcovTay} of $U$ and that of $V$ (i.e.\ the consolidation of \eqref{V0covTay} and \eqref{V1covTay} as explained in Footnote \ref{ftnt:covTay})  and we could then combine these with the expansion \eqref{SigmaN} to arrive at 3-dimensional covariant Taylor expansions for $U$ and $V$ in terms of sums of products of the $\sigma_a$ on the initial surface $\cal C$ to any desired order.  Hence we would be able to obtain a formula for the semiclassical Hadamard subtraction piece $H_\ell$, \eqref{Hadamard}, restricted to our initial surface in terms of these latter expansions.  Moreover, by an extension of this procedure (involving suitable time derivatives of the expansions \eqref{UcovTay}, \eqref{V0covTay},  \eqref{V1covTay} etc.\ as well as the 3-dimensional covariant Taylor expansions for time derivatives of $\Sigma$ that we mentioned above) we could make suitable corresponding definitions for $\partial_t H_\ell$, $\partial_{t'} H_\ell$ and $\partial_t\partial_{t'}H_\ell$.

Thus we would be able to make the definition (cf.\ Definition \ref{Def:YukawaSurfPrelim}) and also the notion of `preliminary surface Hadamard', as discussed for the semiclassical Yukawa model in section \ref{Sect:YukSurf} and for semiclassical QED in section \ref{Sect:ElecSurf}.

\begin{defn}
\label{Def:GravitySurfPrelim}
An initial state $\omega_{\cal C}$ on a constant time surface $\cal C$ equipped with the restrictions to $\cal C$ of the metric $g_{ij}$ and all its time derivatives, $\dot g_{ij}$, $g_{ij}^{(2)}$, $\dots$ is said to be a \emph{preliminary surface Hadamard state for the given values  $g_{ij}, \dot g_{ij}, g_{ij}^{(2)}, \dots $} if, for all $x$ and $x'$ in $\cal C$, the two-point functions $\omega_{\cal C}(\varphi(x)\varphi(x')), \omega_{\cal C}(\pi(x)\varphi(x')), \omega_{\cal C}(\varphi(x)\pi(x')), \omega_{\cal C}(\pi(x)\pi(x'))$  differ from the $H_\ell, \partial_{t} H_\ell, \partial_{t'} H_\ell$ and $\partial_t\partial_{t'}H_\ell$ defined in terms of those $g_{ij}, \dot g_{ij}, g_{ij}^{(2)}, \dots$ respectively by smooth two point functions on $\cal C$.
\end{defn}

However, similarly to the scalar Yukawa case, the problem we face is that only $g_{ij}$ and its first three time derivatives, $\dot g_{ij}, g_{ij}^{(2)}, g_{ij}^{(3)}$, are given to us as part of our initial data.  As in the scalar Yukawa case, we may proceed iteratively to attempt to remedy this lack, now using the equation \eqref{semi-simple} (i.e.\ both the dynamical and the constraint equations) and its successive time derivatives, on our initial surface, $\cal C$, to find the values of the higher time derivatives of $g_{ij}$ restricted to our initial surface.   Again, as a byproduct of this iterative process, we expect to generate a sequence of conditions on the initial state, $\omega_{\cal C}$, of a set of initial data, $g_{ij}, \dot g_{ij}, g_{ij}^{(2)}, g_{ij}^{(3)}, \omega_{\cal C}$ -- and, again, these will be in the form of conditions on the three surface two-point functions, $\omega_{\cal C}(\varphi(x)\varphi(x'))$, $\omega_{\cal C}(\varphi(x)\pi(x'))$  and $\omega_{\cal C}(\pi(x)\pi(x'))$ which must be satisfied for the successive time derivatives of the metric to have finite values.    

The second issue that arises is that, already at the first stage of this iterative process, unlike in the scalar Yukawa model\footnote{We recall here that in the scalar Yukawa model, to obtain the condition on the state in the first stage, i.e.\ the existence of the limit  \eqref{phisqren3d} defining $\omega(\phi^{2 \ \rm{ren}})$, an approximation to $H_\ell$ may be used which depends only on the given initial data, namely the value of $\chi$.} but like in the semiclassical electrodynamics model discussed in Section \ref{Sect:ElecSurf}, the condition on the state in the semiclassical gravity case is that the limit in \eqref{Tabren} exists and one cannot do better than approximate $H_\ell$ by $H_\ell^{\rm{DF}}$ and the latter (see Note (A) after Equation \eqref{DecFolTay} in Section \ref{Sect:star}) depends on the 4th time-derivative of the metric on our initial surface which is already \emph{not} part of our given initial data!   Related to this, the equations \eqref{semi-simple} (with $\omega(T_{ab})$ as in Equation \eqref{semi-hbar}) are not readily expressible in the form (compare Equation \eqref{ddotaEq})
\[
g_{ij}^{(4)} = \hbox{function of the initial state $\omega_{\cal C}$ and $g_{ij}, \dot g_{ij}, g_{ij}^{(2)}, g_{ij}^{(3)}$}
\] 
because $g_{ij}^{(4)}$ occurs both in $\Box R$ and $\Box R_{ab}$ which are directly present in the $1/(4\pi^2)[V_1](x)$ and $\Theta_{ab}(x)$ terms in Equation \eqref{Tabren} as well as (via the $\Box R$ and $\Box R_{ab}$ terms in $H_\ell^{\rm{DF}}$  as explained in Note (A) after Equation \eqref{DecFolTay} in Section \ref{Sect:star}) in the limit in Equation \eqref{Tabren}.  

Nevertheless, there seem grounds for expecting that Equation \eqref{semi-simple} (with $\omega(T_{ab})$ as in Equation \eqref{semi-hbar}) will still, for given $g_{ij}, \dot g_{ij}, g_{ij}^{(2)}, g_{ij}^{(3)}$, both \emph{implicitly} determine which initial states, $\omega_{\cal C}$ (if any) will have finite expectation values for $T_{ab}$ (which in the case of $T_{a0}$ is anyway a prerequisite for being able to say what is meant by the constraint equations holding on $\cal C$) and, for each such state, determine a unique value for $g_{ij}^{(4)}$. 

Motivated by our experience studying that model, we propose the following conjecture:

\begin{conj}
\label{Conj:GravIt}
At the $k$th stage, the $k$th derivative of the semiclassical equations on $\cal C$ will implicitly determine a condition on the initial state $\omega_{\cal C}$ such that, if it can be satisfied, the $k$th time derivative of the stress-energy tensor will exist on $\cal C$ and will also implicitly determine the $k+3$rd time derivative, $g^{(k+3)}$, of the metric in terms of $\omega_{\cal C}$ and the lower order time derivatives.   
\end{conj}

Assuming this to be the case, we are led to a definition of \emph{surface Hadamard state} for semiclassical gravity similar to Definition \ref{Def:YukawaSurf} in the scalar Yukawa case and to Definition \ref{Def:EMSurf} in the case of semiclassical electrodynamics.

\begin{defn}
\label{Def:GravitySurf}
We say that an initial state, $\omega_{\cal C}$, on an initial surface $\cal C$ satisfies
the \emph{surface Hadamard condition} for classical initial data $(g_{ij}, \dot g_{ij}, g^{(2)}_{ij}, g^{(3)}_{ij})$ if (a) it satisfies all the conditions on $\omega_{\cal C}$ that arise at all the stages of the above iterative process starting with the given initial data, and (b) it is preliminary surface Hadamard in the sense of Definition \ref{Def:GravitySurfPrelim} for the full collection of values, $g_{ij}, \dot g_{ij}, g^{(2)}_{ij}, g^{(3)}_{ij}, g^{(4)}_{ij}, \dots $, generated by the above iterative process.
\end{defn}

See the clarifying Remark \ref{Rem:SurfHad} for the relation between the notions of `preliminary surface Hadamard' and `surface Hadamard'.

Parallel to Conjecture \ref{Conj:YukawaIV} for the semiclassical scalar Yukawa model and Conjecture \ref{Conj:EMIV} for semiclassical electrodynamics, we now conjecture:

\begin{conj}
\label{Conj:GravityIV}
Given a 3-manifold, $\cal C$, and any quadruple of classical initial data, $(g^{\rm in}_{ij}, \dot g^{\rm in}_{ij}, g^{(2) {\rm in}}_{ij}, g^{(3) {\rm in}}_{ij})$ on $\cal C$ 
and given an initial state, $\omega_{\cal C}$, on the CCR algebra, $\mathscr{A}({\cal C}, ^3\!g)$, of $\cal C$, which satisfies the appropriate surface Hadamard condition for the given classical initial data, then, 
there will be a unique general solution, $(({\cal M}, g_{ab}), \omega)$ to the semiclassical equations \eqref{semi-simple} (with right hand side defined by Equation \eqref{semi-hbar}), \eqref{KG},  such that $\cal C$ is a Cauchy surface in $\cal M$ and such that the classical gravitational data will arise as the restriction of $g_{ab}$ and its first 3 
time derivatives to $\cal C$ and the state $\omega$ will restrict to $\omega_{\cal C}$, in the sense that $\omega_{\cal C} = \omega\circ {\rm iso}_{\cal C}$.
\end{conj}

Let us remind ourselves here, though, that, as we mentioned in the introduction, since our surface Hadamard condition subsumes the constraint equations, we do not expect that a surface Hadamard state will exist for all quadruples $(g^{\rm in}_{ij}, \dot g^{\rm in}_{ij}, g^{(2) {\rm in}}_{ij}, g^{(3) {\rm in}}_{ij})$.   Furthermore, for quadruples for which a surface Hadamard state exists, we leave the question of uniqueness unanswered.

We may extend the notion of `surface Hadamard' to physical solutions with the definition:
\begin{defn}
\label{Def:PhysSurf}
Given a 3-manifold, $\cal C$, and any pair of classical initial data, $(g^{\rm in}_{ij}, \dot g^{\rm in}_{ij})$ on $\cal C$, we say that a state, $\omega$, is \emph{physical surface Hadamard} for those initial classical data if there exists $g^{(2) {\rm in}}_{ij}, g^{(3) {\rm in}}_{ij}$ such that $\omega_{\cal C}$ is surface Hadamard for  the quadruple $(g^{\rm in}_{ij}, \dot g^{\rm in}_{ij}, g^{(2) {\rm in}}_{ij}, g^{(3) {\rm in}}_{ij})$. 
\end{defn}

With this definition, we are in a position to state the promised clarification of Conjecture \ref{Conj:Phys}:

\begin{conj}
\label{Conj:PhysSurf}
Suppose there exists a physical surface Hadamard state, $\omega_{\cal C}$, for the pair, $(g^{\rm in}_{ij}, \dot g^{\rm in}_{ij})$, of classical initial data on a 3-manifold, $\cal C$, then the $g^{(2) {\rm in}}_{ij}, g^{(3) {\rm in}}_{ij}$ of Definition \ref{Def:PhysSurf} are unique.
\end{conj}

\subsection{\label{Sect:1andhbar} Simplifications for solutions to order $\hbar^0$ and to order $\hbar$}

We may formalize the notions of a \emph{(physical) solution to semiclassical gravity to order $\hbar^0$, respectively $\hbar$}, as consisting of a pair $(({\cal M}, g_{ab}), \omega)$ (all functions of $\hbar$ as well as of spacetime coordinates) where (for each $\hbar$) $({\cal M}, g_{ab})$ is a ($C^\infty$) globally hyperbolic spacetime and $\omega$ a state on the *-algebra $\mathscr{A}({\cal M}, g_{ab})$ such that the equation \eqref{semi-simple} holds to order $\hbar^0$, respectively $\hbar$.   Here we make the assumptions outlined in the last paragraph of Section \ref{Sect:Dephbar} according to which $\omega(\phi(x)\phi(x'))$ takes the form $G_0(x,x') + G_1(x,x')$ where $G_0$ is a smooth two point function of order $\hbar^0$ while $G_1$ differs from the Hadamard subtraction piece, $H_\ell$, of \eqref{Hadamard} by a smooth two-point function.  As explained in Section \ref{Sect:Dephbar}, this entails that the $\omega(T^{\rm ren}_{ab})$ on the right hand side of \eqref{semi-hbar} arises, as in Equation \eqref{semi-refine}, as the sum of a term, $\omega(T_{ab})^{[0]}$ of order $\hbar^0$ and a term, $\omega(T_{ab})^{[1]}$, of order $\hbar$.   In the case of solutions to order $\hbar^0$, we assume that the Hadamard condition plays no r\^ole and we just demand of the state, $\omega$, that, to order $\hbar^0$, its two-point function is smooth; while in the case of solutions to order $\hbar$ we assume that the state satisfies the Hadamard condition but only to order $\hbar$.

As we have already indicated, semiclassical gravity to order $\hbar^0$ would seem to be an approximate theory worthy of further study. On the one hand, it is free from many of the mathematical complications of full semiclassical gravity.  In particular, it does not involve the Hadamard condition or require the inclusion of higher derivative terms.    Instead, as we have already mentioned, the order $\hbar^0$ approximation, \eqref{physsolapprox}, of \eqref{semi-simple} resembles the classical Einstein equation.    On the other hand, it goes beyond classical general relativity in that matter fields are still treated as quantum and also, as we indicated in Section \ref{Sect:Dephbar}, the quantum state, $\omega$, which is being approximated (along with the classical geometry) need not necessarily have the (classical-like) behaviour of a single coherent state, but could, for example, consist of a quantum superposition of coherent states.   (Presumably, one reason that the possibility of this relatively simple approximate theory seems to have been largely overlooked hitherto is that the possibility of states (such as quasifree states with nonvanishing one-point functions) whose two point functions consist of a sum of a term of $O(\hbar)$ and a term of $O(\hbar^0)$   such as those we discuss in Section \ref{Sect:Dephbar} seems to have been neglected in many previous discussions of quantum field theory in curved spacetime and of semiclassical gravity.)

A second reason for being interested in semiclassical gravity to order $\hbar^0$ is that, even if we are ultimately interested in solutions to semiclassical gravity to order $\hbar$, it can surely still serve as a useful first step towards solving a problem of interest at order $\hbar^0$.  Indeed one can adopt a perturbation-theoretic approach and seek a solution to order $\hbar$ as an order $\hbar$ correction to an order 
$\hbar^0$ solution.

For both of the above reasons it is interesting to ask whether there is an initial value formulation for order 
$\hbar^0$ semiclassical gravity.  We make the conjecture (which is in two parts):

\begin{conj}
\label{Conj:hbar0} 
\textbf{(a)}\  Given a 3-manifold, $\cal C$, and any set of ($\hbar$-dependent) classical initial data, $(g^{\rm in}_{ij}, K^{\rm in}_{ij})$ on $\cal C$ and given an initial ($\hbar$-dependent) state, $\omega_{\cal C}$, on the CCR algebra of $\cal C$, whose initial two-point functions, $\omega_{\cal C}(\varphi(x)\varphi(x'))$, 
$\omega_{\cal C}(\pi(x)\varphi(x'))$, $\omega_{\cal C}(\varphi(x)\pi(x'))$, $\omega_{\cal C}(\pi(x)\pi(x'))$, are all smooth to order $\hbar^0$, then, if the constraint equations hold to order $\hbar^0$ (i.e.\ the $0a$ components of Equation \eqref{physsolapprox} hold to order $\hbar^0$) then there will be a (physical) solution, $(({\cal M}, g_{ab}), \omega)$ to \eqref{semi-simple} to order $\hbar^0$, unique to order $\hbar^0$, 
such that $\cal C$ is a Cauchy surface in $\cal M$ and such that, to order $\hbar^0$, the classical gravitational data will equal the restriction of $g_{ab}$ and $\tfrac{1}{2}\dot g_{ab}$ to $\cal C$ and, again to order $\hbar^0$, the state $\omega$ will restrict to $\omega_{\cal C}$.

\smallskip

\textbf{(b)}\  In fact (and this will give us a preferred exemplar for $({\cal M}, g_{ab}, \omega)$ in Part (a)) we conjecture that, given the initial data consisting of $(g^{\rm in}_{ij}, K^{\rm in}_{ij})$ on our 3-manifold $\cal C$ together with the (assumed smooth) order $\hbar^0$ parts of the four bidistributions, $\omega_{\cal C}(\varphi(x)\varphi(x'))$, $\omega_{\cal C}(\pi(x)\varphi(x'))$, $\omega_{\cal C}(\varphi(x)\pi(x'))$, on $\cal C$, then, if the constraint equations hold to order $\hbar^0$, there will be a unique pair consisting of a spacetime, $({\cal M}, g_{ab})$, together with a bisolution, $G_0$, to Equation \eqref{KG} on $({\cal M}, g_{ab})$  such that $\cal C$ is a Cauchy surface in $({\cal M}, g_{ab})$,  $g_{ij}$ restricts to  to $g^{\rm in}_{ij}$ and $\frac{1}{2}\dot g_{ab}$ restricts to $K^{\rm in}_{ij}$ on the initial surface $\cal C$ and such that the (smooth) order $\hbar^0$ parts of $\omega_{\cal C}(\varphi(x)\varphi(x'))$, $\omega_{\cal C}(\varphi(x)\pi(x'))$, $\omega_{\cal C}(\pi(x)\varphi(x'))$, $\omega_{\cal C}(\pi(x)\pi(x'))$ are the restrictions to $\cal C$ respectively of $G^{(0)}(x, x')$ and its $\partial_t$, $\partial_{t'}$ and $\partial_t\partial_{t'}$ derivatives.  And we may then take the full state, $\omega$, in Part (a) above to be $\omega_{\cal C}\circ {\rm iso}_{\cal C}^{-1}$. 

\smallskip

\noindent
(In both (a) and (b) above, it is to be understood that when we refer to a solution to \eqref{physsolapprox} [or, in the case of (a), just of its $0a$ components on the surface $\cal C$] the $\omega(T_{ab})^{[0]}$ on its right hand side is defined to be $\lim_{x' \to x} \mathscr{T}_{ab}(x, x') G_0(x, x')$  where $G_0(x, x')$ is the unique classical bisolution to Equation \eqref{KG} on $({\cal M}, g_{ab})$ with initial data consisting of the smooth order-$\hbar^0$ parts of $\omega_{\cal C}(\varphi(x)\varphi(x'))$, $\omega_{\cal C}(\pi(x)\varphi(x'))$, $\omega_{\cal C}(\varphi(x)\pi(x'))$,  $\omega_{\cal C}(\pi(x)\pi(x'))$, in the sense that the latter are the restrictions to $\cal C$ respectively of $G^{(0)}(x, x')$ and its $\partial_t$, $\partial_{t'}$ and $\partial_t\partial_{t'}$ derivatives.)
\end{conj}

We remark that Conjecture \ref{Conj:hbar0} needs to make explicit mention of the constraint equations since unlike in Conjectures \ref{Conj:GravityIV} and \ref{Conj:PhysSurf}, there is no a surface Hadamard notion which subsumes the constraint equations.

The next (more accurate) level of approximation -- to order $\hbar$ --  is also of great interest since the work of Parker and Simon \cite{Parker:1993dk}. As argued by them, if one expects quantum gravity corrections to appear at order $\hbar^2$, order $\hbar$ semiclassical gravity should be a good approximation to the full quantum theory, at least in some situations.   Also, while not as mathematically simple as semiclassical gravity to order $\hbar^0$, we expect semiclassical gravity to order $\hbar$ to be mathematically simpler than unapproximated semiclassical gravity.  In particular, we expect it to have a simpler initial value formulation due to the fact that the notion of `surface Hadamard' seems to become mathematically simpler when we only require it to order $\hbar$.

Let us first explain this point for the scalar Yukawa model which, we assume can be written
\begin{subequations}
\label{ScalScalarhbar}
\begin{align}
(\partial_a\partial^a - M^2)\chi + \mu = \lambda(\omega^{(0)}(\phi^2) + \omega^{(1)}(\phi^{2\, {\rm ren}}), \label{psieqhbar}\\
(\partial_a\partial^a - m^2 - 2\lambda \chi)\phi =0, \label{phieqhbar}
\end{align}
\end{subequations}
where (cf.\ Section \ref{Sect:Dephbar}) the two point function $\omega^{(0)}(\phi(x)\phi(x'))$ is of order $\hbar^0$ and smooth so that $\omega^{(0)}(\phi^2(x))$ can be defined as its limit as $x\rightarrow x'$, while $\omega^{(1)}(\phi^{2\, {\rm ren}})$ needs to be defined, as in Equation \eqref{phisqren}, by first subtracting the Hadamard subtraction piece, $H_\ell(x,x')$ from the two-point function $\omega^{(1)}(\phi(x)\phi(y))$ before taking the $x\rightarrow x'$ limit, but is of order $\hbar$.   Also, we expect each of our initial quantum states of interest, $\omega_{\cal C}$ on the CCR algebra of a constant time surface, $\cal C$, to arise as $\omega_{\cal C}^{(0)}$ + $\omega_{\cal C}^{(1)}$ where the two-point functions (on $\cal C$), $\omega_{\cal C(0)}^{(0)}\varphi(x)\varphi(y))$, $\omega_{\cal C}^{(0)}(\pi(x)\varphi(y))$, $\omega^{(0)}_{\cal C}(\varphi(x)\pi(y))$, $\omega^{(0)}_{\cal C}(\pi(x)\pi(y))$ will all be smooth while the $\omega_{\cal C(0)}^{(1)}\varphi(x)\varphi(y))$, $\omega_{\cal C}^{(1)}(\pi(x)\varphi(y))$, $\omega^{(1)}_{\cal C}(\varphi(x)\pi(y))$, $\omega^{(1)}_{\cal C}(\pi(x)\pi(y))$ all satisfy the appropriately time-differentiated Hadamard condition to order $\hbar$.

The point is that, in order to define a notion of \emph{surface Hadamard to order $\hbar$} we only need to determine successive time-derivatives of $\chi(t)$ to order $\hbar^0$ since as we pointed out in Section \ref{Sect:Dephbar}, when we restore $\hbar$, the formula for $H_\ell$ into which we need to insert them already has an $\hbar$ prefactor.   So, at the first stage of the iterative process described after Definition \ref{Def:YukawaSurfPrelim}, we may replace  the term $\lambda\omega_{\cal C}(\varphi(x)^{2 \, {\rm ren}})$ in Equation \eqref{ddotaEq} by $\lambda\omega_{\cal C}^{(0)}(\phi(x)^2)$)  and neglect the term $\omega_{\cal C}^{(1)}(\phi^{2\, {\rm ren}})$.  So we may replace Equation \eqref{ddotaEq} by the approximate equation
\[
\ddot \chi  = \nabla^2 \chi  - M^2 \chi  + \mu -  \lambda\omega_{\cal C}^{(0)}(\phi^2) \ \hbox{to order $\hbar^0$}.
\]
Let us also notice that, since we now only solve this equation to order $\hbar^0$, there is no counterpart to the condition that the limit in Equation \eqref{phisqren} needs to exist for us to proceed to the second stage of the iterative process.
Similarly, at the second stage of the iterative process,  we may replace Equation \eqref{3dotaEq} by
\[
\chi^{(3)} = \nabla^2 \dot \chi - M^2 \dot \chi - \lim_{x' \to x} \left( \omega^{(0)}(\pi(x)\varphi(x')) + \omega_{\cal C}^{(0)}(\varphi(x)\pi(x')) \right)\ \hbox{to order $\hbar^0$}
\]
and, again, we need not require the existence of any limits to proceed to the third stage.

Similarly, one may replace Equation \eqref{3rdstage} by
\[
\chi^{(4)} = \nabla^2 \ddot \chi - M^2 \ddot \chi - \lambda \lim_{x' \to x} \left(\omega_{\cal C}^{(0)}(\ddot\varphi(x)\varphi(x')) - \omega^{(1)}(\pi(x)\pi(x'))  - \omega_{\cal C}^{(0)}(\varphi(x)\ddot\varphi(x')\right) \ \hbox{to order $\hbar^0$}.
\]
As when we did things exactly, we may then use Equation \eqref{phieq} to replace $\ddot\phi$ by terms not involving second time derivatives of $\phi$  but we still do not require the existence of any limits to proceed to the fourth stage.   And clearly we will never require the existence of any limits to continue the iterative process indefinitely.

Clearly this results in a much more straightforward procedure (than when we work exactly) by which we may compute order $\hbar^0$ approximations to all the time derivatives of $\chi$ and hence compute approximations to as many terms as we like in the series \eqref{VYukcovTay} that determines the `$V$' term in the Hadamard subtraction piece, $H_\ell$, and hence determines $H_\ell$ itself (given by Equation \eqref{HellYukawa}) on our initial surface $\cal C$.   And similarly we can obtain suitable approximations to order $\hbar$ for the result of acting with the $\partial_t$, $\partial_{t'}$ and $\partial_t\partial_{t'}$ derivatives on $H_\ell$ and then restricting to $\cal C$.   And we may then define a \emph{surface Hadamard state to order $\hbar$} to be one for which, to order $\hbar$, the two point function and its $\partial_t$, $\partial_{t'}$ and $\partial_t\partial_{t'}$ derivatives agree with $H_\ell$ and its corresponding derivatives up to the addition of smooth two-point functions.   

Turning to semiclassical gravity, it is clear that we will similarly have a much more straightforward procedure if we only seek solutions to order $\hbar$ for, similarly to in our scalar Yukawa model, once we are given initial data $(g_{ij}, K_{ij})$ on an initial 3-manifold, $\cal C$, which satisfy the constraint equations (i.e.\ the $0a$ components of \eqref{semi-simple}) in a suitable sense we will then only need to determine the restrictions of successive time derivatives of $g_{ab}$ to an initial surface, $\cal C$, to order $\hbar^0$ in order to determine $H_\ell$ and its $\partial_t$, $\partial_{t'}$ and $\partial_t\partial_{t'}$ derivatives to order $\hbar$ which is all that is needed to define the right hand side of Equation \ref{semi-simple} to order $\hbar$.  (We postpone to later in the discussion the issue of what suitable initial data should consist of and how such data can be found.)  In particular we may avoid the need to consider the successive stages of our iterative process as implicit equations since (just as in the scalar Yukawa model) once we only seek to determine $H_\ell$ to order $\hbar$, we only need determine the successive time derivatives of $g_{ab}$ restricted to our initial surface, $\cal C$, to order $\hbar^0$.   The process of determining these will no longer involve the Hadamard subtraction piece, $H_\ell$ itself, but rather, to order $\hbar^0$, $g^{(2)}_{ij}$ may be obtained by consideration of the order $\hbar^{(0)}$ version, Equation \ref{physsolapprox}, of the semi-classical Einstein equations, \eqref{semi-simple}, while $g^{(3)}_{ij}$ and higher time derivatives of $g_{ab}$ restricted to $\cal C$ may be obtained by consideration of successive time derivatives of Equation \eqref{physsolapprox}. 

Let us remark that, at this point, that this result is consistent with, and gives support to Conjecture \ref{Conj:PhysSurf} holding at least in the case we relax the demand for physical solutions to be exact and we admit approximate solutions to order $\hbar$.

Of course we now also have to concern ourselves with the terms $\frac{1}{4 \pi^2} g_{ab} [V_1](x) + \Theta_{ab}(x)$ in \eqref{Tabren}, which depend on the 4th time derivative of the metric on $\cal C$.  But as we saw in Section \ref{Sect:Dephbar}, when we restore $\hbar$, just as is the case for $H_\ell$, these too acquire $\hbar$ prefactors and so to determine these to order $\hbar$ it also suffices to determine the 4th time derivative of $g_{ab}$ restricted to $\cal C$ to order $\hbar^0$.

To illustrate how we determine  the successive $g^{(n)}_{ij}$ to order $\hbar^0$, let us content ourselves here with the first stage, $n=2$. To order $\hbar^0$, the semiclassical Einstein equations on our initial surface, $\cal C$ will determine $g^{(2)}_{ij}$ since, in Gaussian normal coordinates for a 3-surface, $\cal C$, $G_{ij}$ may be written as $\tfrac{1}{2}\ddot g_{ij} +$ (non-linear) functions of the metric components and their first two spatial derivatives and their first time derivative restricted to $\cal C$.

We must now return to the question, which we postponed, of to what order in $\hbar$ we require the initial data to satsify the constraint equations in order to get started on determining our successive time derivatives of $g_{ab}$ on $\cal C$ to order $\hbar^0$ in order to determine the Hadamard subtraction piece, $H_\ell$ to order $\hbar$.   Ultimately, since we are seeking solutions of semiclassical gravity to order $\hbar$ we will surely need our initial data $(g_{ij}, K_{ij})$ on our initial surface, $\cal C$, to satisfy the constraint equations to order $\hbar$ and the right hand side of these (see Equation \eqref{GravConstraints}) will already require us to know the Hadamard subtraction piece, $H_\ell$ to order $\hbar$.   Thus it might seem that we cannot fully avoid the need to get involved with solving one of the implicit equations that we needed to get involved with when we attempt to solve semiclassical gravity exactly.  However, if we take the perturbative approach that we mentioned above as our second reason for studying semiclassical gravity to order $\hbar^0$, then we would expect it to be possible to avoid that.  We envisage a strategy in which one first solves both the constraint equations and the dynamical equations to order $\hbar^0$.   Then the solution to the constraint equations to order $\hbar^0$ will suffice to determine the all the time derivatives of $g_{ab}$ on $\cal C$ to order $\hbar^0$ and this will suffice to determine $H_\ell$ to order $\hbar$.   Then, equipped with that order $\hbar$ $H_\ell$, one can revisit the solution to the constraint equations and, with a perturbative approach, adjust it to hold to order $\hbar$ at the same time as using a perturbative approach to find the solution to the dynamical equations to order $\hbar$.   We will not attempt to spell out the details further in the present paper but we do think what we have written above constitutes a \emph{prima facie} case that an initial value conjecture similar to Conjecture \ref{Conj:PhysSurf} should at least hold with the appropriate insertions of the phrase `to order $\hbar$'.  

\section{Semiclassical gravity in the presence of quantum state collapses and the `What jumps?' question}
\label{Sect:collapse}

As discussed in the introduction, we wish to extend the notion of a physical solution to semiclassical gravity to allow for quantum state collapses to occur, according to some stochastic rule, on certain random (but non-intersecting) Cauchy surfaces, which we shall call \emph{collapse surfaces}.  A semiclassical solution with collapses will then consist of a sequence of \emph{epochs} each of which may be identified with the restriction to the region between two (non-intersecting) Cauchy surfaces in a semiclassical solution without collapses.   And we shall always assume that, except at its initial and final Cauchy surfaces, each epoch is in fact a (smooth) physical solution in the sense we discussed in Section \ref{Sect:SSC}.  
    
Successive epochs are then glued together so that the final Cauchy surface of each epoch is identified, as a manifold, with the initial Cauchy surface of the subsequent epoch.\footnote{Obviously this gluing assumption entails that all epochs have the same topology.}   We shall not be concerned here with the full stochastic rule that determines when collapses happen and how the quantum state changes when they do.  Instead we shall focus just on one aspect of the rule.  Namely, we shall ask what can be said, independently of other details about the rule, about the changes, or \emph{jumps}, in the classical data, $(g_{ij}, \dot g_{ij})$ on a collapse surface which (as a result of the sudden change in the right hand side of the constraint equations) we expect will necessarily accompany the sudden changes in the quantum state on the collapse surfaces.  We will arrive at a tentative partial solution to this question, and will attempt to make clear the open questions that remain.

To study this question, we shall rely on the validity of our initial value conjecture, Conjecture 
\ref{Conj:PhysSurf} in the case of exact physical solutions.   However, everything we say below will also apply equally in the context of the order zero or 1 in $\hbar$ approximate solutions which were discussed in Section \ref{Sect:1andhbar} provided we use the appropriate replacements for Conjecture \ref{Conj:GravityIV}) as discussed in that subsection. 

Let us denote the metric in the $n$th epoch by $g^n_{ab}$ and the quantum Hadamard state by $\omega^n$.  For a physical semiclassical solution with collapses to take the form described above, it must be that the (unknown, stochastic) collapse rule will ensure that, just as $\omega^n$ will be a Hadamard state with respect to the metric $g^n_{ij}$ in the $n$th epoch,  the post-collapse state, $\omega^{n+1}$, will also be a Hadamard state with respect to the metric $g^{n+1}_{ij}$ in the $n+1$th epoch.  We expect that this will come about dynamically in the way we next describe:   

We shall assume from the outset that the volume element $\sqrt{{\rm det}^3g}\,d^3x$ (see after Equation \eqref{CCR}) will remain unchanged during a collapse, so the CCR algebra of a collapse surface is well defined. This is guaranteed by requiring that the metric be continuous on the collapse surface.

Just before the quantum state-collapse occurs at the end of the $n$th epoch, we can restrict the metric and its first derivative to the final Cauchy surface, ${\cal C}^{n {\rm fi}}$, of that epoch, obtaining, say the classical data, $(g^{n   {\rm fi}}_{ij}, \dot g^{n  {\rm fi}}_{ij})$, and we may also consider the restriction, $\omega^n_{{\cal C}^{n   {\rm fi}}}$ of $\omega^n$ to that surface, which will clearly necessarily be a physical surface Hadamard state.  (Strictly we mean $\omega^n\circ {\rm iso}_{{\cal C}^{n   {\rm fi}}}$ where ${\rm iso}_{{\cal C}^{n  {\rm fi}}}$ is defined as in Equation \eqref{iso} for the $n$th epoch).  $\omega^n_{{\cal C}^{n {\rm fi}}}$ will change abruptly to a new state, $\omega^{n+1}_{{\cal C}^{n+1  {\rm in}}}$ on the CCR algebra of the same surface, now thought of as the initial surface,  ${\cal C}^{n+1  {\rm in}}$, of the $n+1$st epoch and at the same time, the classical data will change to new classical data, $(g^{n+1  {\rm in}}_{ij}, \dot g^{n+1  {\rm in}}_{ij})$ and it must be that our collapse rule is such as to guarantee that $\omega^{n+1}_{{\cal C}^{n+1 {\rm in}}}$ will be a physical surface Hadamard state for the latter data. 

It will follow by the initial value conjecture 
\ref{Conj:PhysSurf} that we will have a well-defined physical semiclassical solution (i.e.\ spacetime together with Hadamard quantum state satisfying our smoothness condition in $\hbar$) to the future of ${\cal C}^{n+1  {\rm in}}$ which will be valid until the same collapse rule causes a further state collapse to occur on some Cauchy surface to the future, thus demarcating the end of the $n+1$st epoch and the beginning of the $n+2$nd epoch and so on.   So we will have a well-defined time-evolution (involving our stochastic collapse rule) that takes the solution forwards in time from one epoch to the next.

Clearly, studying the possible answers to the question of `what jumps' for the classical data will be an important first step towards understanding the possible options for the full collapse rule.   Before we discuss this question for semiclassical gravity, it will be helpful to first discuss it for our scalar Yukawa model and for semiclassical electrodynamics.

In both these models, we assume that the collapses take place on certain random constant time surfaces for some fixed Lorentz frame in Minkowski space.  And we take the epochs to consist of solutions in the sense of Section \ref{Sect:SSC} in the regions bounded by successive such surfaces.

\subsection{Scalar Yukawa model with quantum state collapses}

Suppose our unknown collapse rule causes a Hadamard state, $\omega^-$ in one epoch to change to a Hadamard state $\omega^+$  in the subsequent epoch. The question we wish to address is how do we expect the classical data, i.e.\ $\chi $ and $\dot  \chi$, to change on the collapse surface, $\cal C$ at the junction of the two epochs.   We propose that the right prescription is that \textit{nothing} happens to $\chi$ and $\dot \chi$ -- i.e.\ they are continuous.   

With this assumption, provided the stochastic collapse rule causes a surface Hadamard state, $\omega^-$ for $\chi$ and $\dot \chi$ on the collapse surface, $\cal C$ to suddenly change to to another surface Hadamard state, $\omega^+$ for the same classical data, then the solution to the future of $\cal C$ will (until the next state-collapse) simply be the solution whose initial data on $\cal C$ is $(\chi, \dot \chi, \omega^+)$.

Note that, in view of the sudden change in the state, we cannot regard the semiclassical equations \eqref{ScalScalar}, as holding in any neighbourhood that includes any part of the collapse surface.  Nevertheless, we expect our above continuity-of-classical Cauchy-data rule to still allow us to determine the solution beyond the collapse surface.

\subsection{Semiclassical scalar electrodynamics with quantum state collapses}

We now turn to the same question for our scalar electrodynamics model of Equations \eqref{SemiElectro}, \eqref{Maxwell}, \eqref{ScalElecCoul}: Suppose our unknown collapse rule causes a Hadamard state, $\omega^-$ in one epoch to change to a Hadamard state $\omega^+$  in the subsequent epoch.
One might naively think that one could do similarly to in the scalar Yukawa model and set $\bf E^+$ and ${\bf B}^+$ (i.e.\ the values of $\bf E$ and $\bf B$ just to the future of the collapse surface) equal to ${\bf E}^-$ and ${\bf B}^-$ -- i.e.\ the values of $\bf E$ and $\bf B$ just to the past of the collapse surface.  But this cannot work because $\omega^-(\rho)$  will not (in general) equal $\omega^+(\rho)$ and $\bf E$ needs to satisfy the Gauss constraint (i.e.\ \eqref{Max1}) both before and after the collapse and hence, at least generically, must jump too.  However, there is a natural modification of that prescription where we make what would seem to be the \textit{minimal} jump in $\bf E$ and $\bf B$ compatible with the Gauss constraint.   Namely, we assume that $\bf B$ does not jump and that, if we do the Helmholtz decomposition on ${\bf E}^-$ (see Appendix \ref{app:Y})
\begin{equation}
\label{EjumpA}
{\bf E}^- = -{\bf\nabla}\Theta^- + {\bf Q}
\end{equation}
where $-{\bf\nabla}.{\bf\nabla}\Theta^-=4\pi\rho^-$ and ${\bf\nabla}.{\bf Q}=0$,
then 
\begin{equation}
\label{EjumpB}
{\bf E}^+ = -{\bf\nabla}\Theta^+ + {\bf Q}
\end{equation}
where $-{\bf\nabla}.{\bf\nabla}\Theta^+ = 4\pi\rho^+$.   (With solution for $\Theta^-({\bf x})$ and $\Theta^+({\bf x})$ in terms of $\rho^-$ and $\rho^+$ respectively given by $\int \frac{\rho^\pm({\bf x}')}{|{\bf x} - {\bf x}'|} d^3x'$. )

Next notice that, if we work in Coulomb gauge, then we may identify the $\Theta$ and $\bf Q$ of the Helmholtz decomposition of $\bf E$ with the electrical potential, 
$\Phi$ and $\frac{\partial {\bf A}}{\partial t}$ respectively.
In fact, the prescription \eqref{EjumpA}, \eqref{EjumpB} is equivalent to saying that the Coulomb gauge version of the equations \eqref{ScalElecCoul} holds at all times except the collapse time, together with the prescription that the electric potential, $\Phi$, jumps so that, $\Phi^\pm$, is always the Coulomb potential for $\omega^\pm(\rho)$ while the magnetic vector potential, $\bf A$, is continuous.

Note that it is an immediate consequence of \eqref{ScalElecCoul} that -- except at the collapse time, when the equations do not hold and anyway $\partial\rho/\partial t$ will not generically exist -- the continuity equation ${\bf \nabla}.{\bf j} + \partial\rho/\partial t =0$ will hold automatically. (But the continuity equation will not hold or even make any sense at the collapse time.)

Lastly, and in preparation for our discussion of the semiclassical gravity case, let us notice that, physically, our prescription amounts to saying that the part of the electric field that jumps is the (nonpropagating, longitudinal), part that is tied, via the Gauss' law constraint, \eqref{Max1} to the expectation value of the charge density, $\rho$, while the (radiative, transverse) part of the electric field (together with the entire magnetic field) remains continuous and is, so to speak, unaffected by the quantum state collapse.

\subsection{Semiclassical gravity with quantum state collapses}

We now turn to discuss the counterpart to all the above for physical solutions to the equations \eqref{semi-simple} and \eqref{KG}.  

Suppose, again, that our unknown collapse rule causes a Hadamard state, $\omega^-$, in one epoch to change to a Hadamard state, $\omega^+$,  in the subsequent epoch, where the epochs are glued together at a common (curved) Cauchy surface.

Just as in the electromagnetic case, where the expectation value of the electrical charge density, $\rho$, on the right hand side of the Gauss' law constraint equation, \eqref{Max1} (equivalently \eqref{Coulomb1})  undergoes a sudden change, so, in the gravitational case, the, now four, quantities $\omega(T^{{(\ell)}\rm ren\, 00})$, $\omega({T^{(\ell){\rm ren,0}}_{\ \ \ \ \ \ \ \ \, i}})$ on the right hand side of the constraint equations, \eqref{GravConstraints} will undergo sudden changes.  And, as clearly explained in \cite{Page}, because of such sudden changes, the semiclassical Einstein equations will not hold across the collapse surface and also the Bianchi identity will fail -- in analogy to what we saw above about the failure of current conservation in semiclassical electrodynamics with collapses.  The question we are interested in is: \textit{What pieces of the induced metric, $g_{ij}$, and extrinsic curvature, $K_{ij}$ ($=\tfrac{1}{2}\dot g_{ij}$), jump (i.e.\ fail to be continuous) and what is our prescription for how they change, or `jump'?}  

We shall discuss one possible tentative solution to this question where we assume at the outset that $g_{ij}$ does not jump.  We remark concerning this assumption that, in any case, as we mentioned above, we require the volume element, $\sqrt{{\rm det}^3g}\,d^3x$ not to jump in order to have a well defined CCR algebra on our collapse surface and this requirement will of course be automatically satisfied if $g_{ij}$ does not jump.   With this assumption, we presumably need to decompose $K_{ij}$ into two suitable pieces -- one piece, which jumps, having four independent components and the remaining piece, which does not jump, having two independent components.   In analogy with semiclassical electrodynamics, to identify what these pieces are, we consider the rank-2 tensor version of the Helmholtz decomposition, displayed in Equation \eqref{DanHoltz} of Appendix \ref{app:Y}, applied to $K_{ij}$,
\begin{equation}
\label{KanHoltz}
K_{ij}=\left(D_i D_j-\frac{1}{3}g_{ij}D^l D_l\right)\alpha + g_{ij}\beta+D_{(i}C_{j)}+L_{ij}
\end{equation}
where $C_i$ is divergence free and $L_{ij}$ transverse traceless and $D$ denotes a covariant derivative compatible with the induced $3$-geometry on the collapse Cauchy surface.  Note by the way that we see from
\eqref{KanHoltz} that $K (=K_a^{\;\; a}) =3\beta$.

Our tentative proposal is that the term in this decomposition that \emph{does not} jump is the transverse traceless term, $L_{ij}$.\footnote{We remark that, with this prescription, $K$ (the trace, $K_i^{\;\; i}$, of $K_{ij}$) will jump and therefore, if we wished to have a global field *-algebra for more than a single epoch we would have to deal with a Klein-Gordon equation (which is easily shown to take the form $\frac{\partial^2\phi}{\partial t^2} + K\frac{\partial\phi}{\partial t} - D_iD^i\phi=0$ where $D_i$ is the covariant derivative in the surface at constant time $t$ with respect to $g_{ij}$) which has a coefficient, $K$, which is discontinuous!  But it seems that, for our purposes we can content ourselves with having a separate field algebra for each epoch so the issue does not arise.}   

Two pieces of evidence that this is a reasonable proposal are (a) that it has the right number of independent components, i.e.\ 2.   In fact we make the conjecture:

\begin{conj} (In the presence of a continuous metric $g_{ij}$) the 4 constraint equations after a quantum state collapse are just sufficient to uniquely determine the remaining part of $K_{ij}$ postcollapse (i.e.\ the part of Equation \eqref{KanHoltz} involving  $\alpha$, $\beta$ and $C_j$ which together have 4 independent components) in terms of the values of $g_{ij}$ and $K_{ij}$ pre-collapse and the changes in $\omega(T^{{(\ell)}\rm ren\, 00})$, $\omega^({T^{(\ell){\rm ren,0}}_{\ \ \ \ \ \ \ \ \, i}})$ when $\omega$ changes from $\omega^-$ to $\omega^+$.  
\end{conj}

Secondly the transverse traceless part of $K_{ij}$ (together with that of $g_{ij}$) represents the part of the gravitational field which is a gravitational wave and not `tied' to the sources $\omega^-(T^{{(\ell)}\rm ren\, 00})$, $\omega^-({T^{(\ell){\rm ren,0}}_{\ \ \ \ \ \ \ \ \, i}})$, in analogy with the way that, in the electromagnetic case, we argued that the part of the electric field which remains continuous is the transverse part of the electric field which (together with the transverse part of the magnetic field) represents an electromagnetic wave which is not tied to the electrical charge density.

Assuming the validity of the conjectures we made on the initial value problem in Section \ref{Sect:SSC}  about physical solutions (or, if we work to order $\hbar$ or order $\hbar^0$, the conjectures we made in Subsection \ref{Sect:1andhbar} about the initial value problem for those approximate theories) and assuming the above jump rule, we can now adumbrate a well-defined answer to the question ``\emph{What is a solution to the initial value problem for semiclassical gravity with quantum state collapses?}''.  
 
Let us assume given gravitational data $(g_{ij}, K_{ij})$ on an initial Cauchy surface
together with given initial data for a surface Hadamard state on the CCR algebra of that surface for those gravitational data as defined in Definition \ref{Def:GravitySurf}.\footnote{It is irrelevant whether the starting surface itself is a collapse surface, provided that, if it is not, we are referring to the data after a quantum state collapse and its associated jump in the classical data.}    By Conjecture \ref{Conj:PhysSurf} there will be a unique physical solution to semiclassical gravity without collapses to the future of our initial surface.   We take the physical solution in the presence of collapses to coincide with that solution up to the first collapse surface to its future.  On that collapse surface, we assume that the data for the quantum state will jump according to a stochastic collapse rule, which has the feature that if the classical Cauchy data  $(g_{ij}, K_{ij})$ jumps according to the above jump rule, then the data for the new quantum state will satisfy the physical surface Hadamard condition for the new $(g_{ij}, K_{ij})$.  What such a collapse rule could be and how it will ensure that feature we shall not discuss further here.

With that assumption, and after adjusting the classical Cauchy data according to our jump rule, again by Conjecture 
\ref{Conj:PhysSurf} we will have a semiclassical solution up until the next collapse surface -- i.e.\ in the next epoch -- and so on.   So provided there is a collapse rule that has the above feature, we will clearly have a well-defined unique prescription for the entire future time evolution of our initial data in the presence of collapses.    The prescription will of course inherit a stochastic nature from the stochastic nature of the jumps due to the collapse rule on each of the collapse surfaces.

Moreover, let us notice that we may dispense with the above assumption if we work to order 
$\hbar^0$ since, as we explained in Section \ref{Sect:1andhbar}, there is then no surface Hadamard condition that needs to be satisfied.  Thus at that level of approximation, there is no restriction on any stochastic collapse rule and we may apply any such rule and pass to a solution in the next epoch after adjusting the classical initial data on our collapse surface according to our jump rule.  

An early exploration of ideas, similar in spirt  to those developed in this work, was carried out using  a perturbative approach in \cite{DiezTejedor:2011ci} and \cite{Canate:2018wtx} in the context of inflationary cosmology.  There, the high level of symmetry of the background, and the use of coherent  states helped simplify the problem immensely.   Those works  studied the transition as a result of a spontaneous collapse on a single collapse surface (assumed to be at some constant cosmological time)  from  an initial state with an initial exact FLRW symmetry to a state in which that symmetry is broken.  The  relationship  with the present  work is however not truly straightforward, as the small parameter in the perturbative treatment employed there was the magnitude of the  metric  inhomogeneity, rather than the kind of expansion in  $\hbar$ often contemplated in this work.  However, specializing our formalism and our jump rule here to that case, and making the appropriate adjustments to  ensure the perturbative expansions  coincide, the general approach developed in this work should reproduce the results of those papers.

\section{Discussion and  conclusions}
\label{Sect:conc}

In this paper we have been concerned with understanding two aspects of semiclassical gravity. The first was the initial value problem for semiclassical gravity and here we made a number of conjectures about the form its solution will take. The second has to do with attaching to the theory the 
possibility of `quantum state collapse' in a sensible way.

In both cases we have taken lessons from two comparatively simpler problems.  Namely, a semiclassical Yukawa model of two scalar fields, and semiclassical electrodynamics, in which a quantum field interacts with the classical electromagnetic field. In doing so, we have put 
forward a number of conjectures about what it means for the initial value formulation of a semiclassical theory to be well posed.
 
One lesson that we have learned is that the notion of `Hadamard initial data' encoded here with the notion \emph{surface Hadamard state} imposes an infinite tower of constraints on initial data that can be obtained by successive differentiation and application of the field equations on the initial surface. Concerning the notion of state collapse, we have understood that the constraints of a semiclassical theory -- which occur both in semiclassical gravity and electrodynamics -- impose conditions on `what jumps' in the classical configurations upon quantum state collapse, assuming that collapses 
occur on certain Cauchy surfaces that we call \emph{collapse surfaces}. We have remained agnostic about what exact mechanism could cause state collapse, but our discussion of our jump rule is independent of that.

Another feature of our paper is that our formalism is sufficiently general as to allow explicitly for the possiblility (which is sometimes neglected in other work on semiclassical gravity) of quantum states for which the two-point function contains a term of order $\hbar^0$ as well as a term of order $\hbar$ (such as is the case e.g.\ for a quasifree state with a non-vanishing one-point function). This is because, as we pointed out in the introduction (see especially Footnote \ref{FRLWBirk}), in an approach to semiclassical gravity in which the only classical field is the gravitational field and all other fields are regarded as quantum, then it appears essential, to do cosmology, to consider such states.

Throughout this paper we have also adapted the viewpoint of Parker and Simon \cite{Parker:1993dk} (originally proposed to eliminate late-time runaway solutions) by defining \emph{physical} semiclassical solutions to be those that depend smoothly on $\hbar$ at $\hbar = 0$. We have furthermore conjectured that the space of physical solutions of semiclassical gravity, thus defined, can be characterized by only two pieces of classical initial data -- namely the metric and the extrinsic curvature --  with the remaining two pieces of (higher order) classical data being determined from the first two pieces by imposing the `physicality' condition. Such a conjecture was motivated by studying simpler problems in the context of ordinary differential equations. 

Furthermore on this note, we have emphasized the importance of an approximate theory intermediate between full semiclassical gravity and classical general relativity -- namely semiclassical gravity to order $\hbar^0$ -- which we feel deserves attention in its own right as a distinct theory from classical general relativity.  See especially Conjecture \ref{Conj:hbar0} in Section \ref{Sect:1andhbar}. The reason for this, as we have argued, is that the $O(\hbar^0)$ part of the expectation value of the stress-energy tensor in a quantum state of the sort mentioned above will not necessarily behave like a classical stress-energy tensor, as is the case for semiclassical gravity with a quantum field in a single coherent state (i.e.\ a single quasi-free state with nonvanishing one-point function).  Indeed, we have seen that superpositions of such coherent states have a non-classical and non-trivial contribution at zeroth order in $\hbar$ (with e.g.\ a dependence on $\hbar$ that goes like $A\exp(-B/\hbar)$).
  
We also discussed semiclassical gravity to order $\hbar$, which had also been discussed by Parker and Simon in \cite{Parker:1993dk} and which is intermediate between the $O(\hbar^0)$ theory and the theory of the physical solutions of full semiclassical gravity.

In any case, we have put forward a number of conjectures for the semiclassical Yukawa, electrodynamics and gravity theories that we hope will stimulate mathematical work to prove (or disprove) them.  And we hope that the conjectures themselves will already serve to promote a better understanding of the structural properties of semiclassical gravity -- and of other semiclassical theories.

We find it important to stress the importance of the Cavendish experiment of Page and Geilker \cite{Page}, which has had a big impact on our understanding of the status of semiclassical gravity as an effective theory. As we have argued in the introduction to this paper we find that some of their conclusions concerning the interpretation of state collapses were premature.   We feel that we have, instead, made at least a \emph{prima facie} case that a theory of semiclassical gravity with quantum state collapses is possible, and offers an alternative way to explain the results of their Cavendish experiment without the need to invoke full quantum gravity. This does not mean that we expect semiclassical gravity with collapses to be a fundamental theory, but indeed we endorse the view that it should arise as an effective limit of some yet-unknown quantum gravity theory.

\section*{Acknowledgements}
B.A.J-A. is supported by a CONACYT postdoctoral fellowship. T.M. is supported by a CONACYT PhD scholarship. B.A.J.-A. and D.S. acknowledges partial financial support from  PAPIIT-UNAM, Mexico (Grant No.IG100120 ) and CONACYT project 140630. D.S. also thanks the Foundational Questions Institute (Grant No. FQXi-MGB-1928) and the Fetzer Franklin Fund, a donor advised by the Silicon Valley Community Foundation. 

\appendix

\section{Helmholtz Theorem for Vector Fields and its Generalization to Symmetric 2-Tensors}
\label{app:Y}

Let $(M, h_{ij})$ be a 3-dimensional Riemannian manifold.

Let $\bf F$ be a vector field on $(M, h_{ij})$.  Then \textit{the Helmholtz decomposition theorem} says that we can find a scalar field $\Theta$ and a vector field $\bf Q$ with $\bf\nabla.{\bf Q}=0$ such that
\begin{equation}
\label{Helmholtz}
{\bf F} =-{\bf\nabla}\Theta + {\bf Q}
\end{equation}
Note that in ${\mathbb R}^3$, any ${\bf Q}$ with $\bf\nabla.{\bf Q}=0$ arises in the form
${\bf Q} = {\bf\nabla}\times {\bf R}$ for some vector field $\bf R$.

In index notation, we have $F_i=-\nabla_i\Theta + Q_i$ with $\nabla^iQ_i=0$.

There are two counterparts to this result for a symmetric 2-tensor, say $S_{ij}$, on $(M, h_{ij})$ \cite{York}:
\begin{equation}
\label{York1}
S^{ij}=S_t^{ij}+\nabla^iX^j+\nabla^jX^i \quad \hbox{where} \ \ \nabla_iS_t^{ij}=0,
\end{equation}
and
\begin{equation}
\label{York2}
S^{ij} =S_{TT}^{ij} + \nabla^iY^j+\nabla^jY^i-\frac{2}{3}h^{ij}\nabla_kY^k+\frac{1}{3}Sh^{ij}
\end{equation}
where $\nabla_iS_{TT}^{ij}=0$, ${S_{TT}}^i_{\;\; i}=0$ and $S$ means the trace ($S_i^{\;\; i}$).

Combining the latter with Helmholtz, \eqref{Helmholtz}, we easily obtain  
\begin{equation}
\label{DanHoltz}
S_{ij}=\left(\nabla_i\nabla_j-\frac{1}{3}h_{ij}\nabla^l\nabla_l\right)\alpha + h_{ij}\beta+\nabla_{(i}C_{j)}+L_{ij}
\end{equation}
where $\alpha$ and $\beta$ are scalar fields (so each count as one independent component), $C^i$ is a vector field with $\nabla_iC^i=0$ (so counts as 3-1=2 independent components) and $L_{ij}$ is a symmetric tensor which is transverse and traceless, i.e.\ satisfies $\nabla^iL_{ij}=0$ and $L^i_i=0$ (so counts as 6-3-1=2   independent components).  Note that the sum of the independent components 1+1+2+2=6 checks as matching the number of independent components of $S_{ij}$.

\end{document}